\documentclass[aps,prl,twocolumn,superscriptaddress,longbibliography,nobalancelastpage]{revtex4-2}

\usepackage{lipsum}
\usepackage{graphicx}
\usepackage{amsmath}
\usepackage{amssymb}
\usepackage{amsthm}
\usepackage{color}
\usepackage{enumitem}
\usepackage[normalem]{ulem}
\usepackage[utf8]{inputenc}
\usepackage{environ}
\usepackage[svgnames]{xcolor}
\usepackage[colorlinks = true,allcolors = {Blue}]{hyperref}
\usepackage{orcidlink}
\usepackage{physics}
\usepackage{soul}

\usepackage[caption=false]{subfig}
\newcommand{\phantomsubfloat}[1]{{\captionsetup[subfigure]{labelformat=empty}\subfloat{#1}}}

\newcommand{\id}[0]{{1\!\!1}}
\newcommand{\dt}[0]{\ensuremath{\delta t}}
\newcommand{\MPS}[1][]{\ensuremath{\ket{\mathtt{MPS}#1}}}
\newcommand{\e}[1]{e^{#1}}

\newtheorem{theorem}{Theorem}

\begin{document}

\title{Disentangling magic states with classically simulable quantum circuits}
\author{Gerald E. Fux\,\orcidlink{0000-0002-7912-0501}}
\affiliation{The Abdus Salam International Center for Theoretical Physics (ICTP), Strada Costiera 11, 34151 Trieste, Italy}
\author{Benjamin B\'eri\,\orcidlink{0000-0001-9933-9108}}
\affiliation{T.C.M. Group, Cavendish Laboratory, University of Cambridge, J.J. Thomson Avenue, Cambridge, CB3 0HE, UK\looseness=-1}
\affiliation{DAMTP, University of Cambridge, Wilberforce Road, Cambridge, CB3 0WA, UK}
\author{Rosario Fazio\,\orcidlink{0000-0002-7793-179X}}
\author{Emanuele Tirrito\,\orcidlink{0000-0001-7067-1203}}
\affiliation{The Abdus Salam International Center for Theoretical Physics (ICTP), Strada Costiera 11, 34151 Trieste, Italy}
\affiliation{Dipartimento di Fisica ``E. Pancini", Universit\`a di Napoli ``Federico II'', Monte S. Angelo, 80126 Napoli, Italy}
\date{\today}

\begin{abstract}
We show that states obtained from deep random Clifford circuits doped with non-Clifford phase gates (including T-gates and $\sqrt{\mathrm{T}}$-gates) can be disentangled completely, provided the number of non-Clifford gates is smaller or approximately equal to the number of qubits.
This implies that Pauli expectation values of such states can be efficiently simulated classically, despite them exhibiting both extensive entanglement and extensive nonstabilizerness.
We prove this result analytically using a quantum error correction formulation, demonstrate its applicability numerically, and discuss consequences for the disentanglability of states generated through Hamiltonian dynamics.
We show that this result implies a novel representation of approximate state designs that can also facilitate their efficient generation, and we propose a novel quantum circuit compression scheme for Clifford circuits doped with non-Clifford phase gates.
\end{abstract}

\maketitle


Many-body quantum states are generically hard to simulate classically due to their Hilbert space growing exponentially with system size.
There are, however, several classes of states that can be simulated classically with only polynomial resources.
One such class is the class of tensor network states, particularly matrix product states (MPS)~\cite{schollwock2011thedensitymatrix, orus2019tensor, ran2020tensor}, which are efficient when entanglement does not grow with system size~\cite{vidal2002efficient,vidal2004efficient,eisert2010colloquium}.
Another class of states called \emph{stabilizer states} may exhibit high entanglement but remain efficiently classically simulable by the Gottesman-Knill theorem, establishing the \emph{stabilizer formalism}~\cite{gottesman2004stabilizer,  gottesman1998theoryoffaulttolerant, gottesmann1998faulttolerant, aaronson2004improvedsimulation}.
Stabilizer states are generated by arbitrary combinations of the Hadamard, $\pi/4$ phase, and controlled-NOT gates applied to a computational basis state.
This gate set generates the Clifford group, which with just one additional non-Clifford gate, such as the T-gate ($\pi/8$ phase gate), becomes a universal gate set.

\begin{figure}
	\phantomsubfloat{\label{fig:clifford-circuit}}%
	\phantomsubfloat{\label{fig:theorem}}%
	\centering
	\includegraphics[width=1\linewidth]{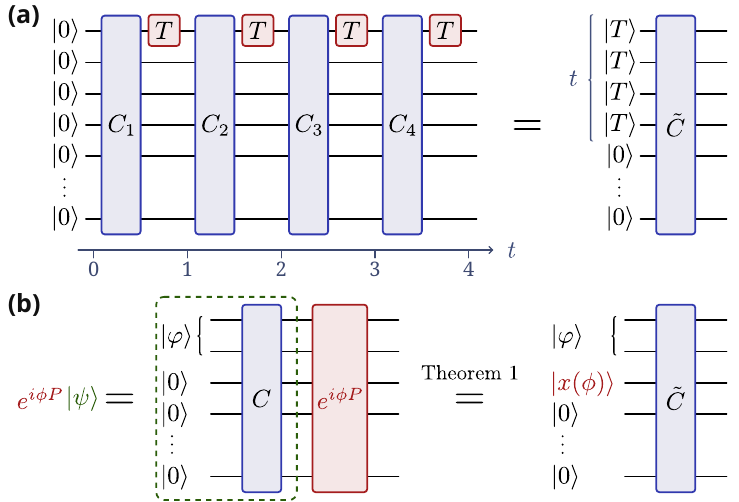} 
	\caption{\label{fig:circuits}%
		Panel~(a): For $t\lesssim N-O(1)$, the action on $\ket{0}^{\otimes N}$ of a quantum circuit, with a uniformly random global Clifford $C_i$ (i.e., deep random Clifford circuit, blue boxes) on all qubits followed by a  T-gate (red boxes) at every time step equals (up to a phase) the state on the right (with $\tilde{C}$ Clifford) in almost all instances for large $N$.
		Panel~(b): a general phase gate $\e{i\phi P}$ applied onto  the logical state $\ket{\varphi}$ encoded by a Clifford unitary $C$. When the QEC conditions for $P$ hold, Theorem~\ref{th:theorem} shows that this equals the logical state $\ket{\varphi}\otimes \ket{x(\phi)}$ encoded by a Clifford unitary $\tilde{C}$.
	}
\end{figure}

The amount of non-Clifford resources necessary to create a state is called \emph{nonstabilizerness}, or \emph{magic}~\cite{bravyi2005universal, veitch2014resource,chitambar2019quantum,bravyi2016trading}.
Nonstabilizerness has been rigorously defined using quantum resource theories, with various measures proposed to quantify it~\cite{gross2021schur, leone2022stabilizerrenyientropy, haug2023scalable, haug2023quantifying, haug2023stabilizer, leone2024stabilizer,turkeshi2024magic}, and it has emerged as a key factor for the classical simulation hardness of quantum systems~\cite{yoganathan2019quantum,white2021conformal,lami2023nonstabilizerness, chen2024magic, tarabunga2023manybodymagic, tarabunga2024nonstabilizerness,tarabunga2024magic,tarabunga2024critical,bejan2024dynamical,bluvstein2024logical, oliviero2022measuring, haug2024efficient, niroula2024phase}.
While states with low entanglement or low nonstabilizerness can be simulated efficiently, states with both a high entanglement \emph{and} nonstabilizerness cannot be simulated using the stabilizer formalism or tensor network methods.
A prominent example of quantum circuits that produce both high entanglement and nonstabilizerness are Clifford circuits interspersed with T-gates, as shown in Fig.~\ref{fig:clifford-circuit}.
Such Clifford + T circuits are universal and have been shown to $\epsilon$-approximate unitary $k$-designs when choosing the Clifford circuits at random for $t=O[k^4 \log^2(k) \log(1/\epsilon)]$ layers~\cite{haferkamp2023efficient}.

If we apply such Clifford + T circuits on a computational basis initial state [Fig.~\ref{fig:clifford-circuit}], is there a critical $t$ beyond which entanglement and magic become inextricable?
We show that it is typically possible to cast this evolution into a single Clifford circuit applied to a magic product state, even if the number $t$ of non-Clifford gates is of the order of the number $N$ of qubits.
Concretely, we find that disentangling works for $t \lesssim N - 1.607$ on average.
As a consequence, one can classically simulate Pauli expectation values for such evolutions efficiently in almost all instances.
We prove our result analytically and provide exact conditions in terms of quantum error correction (QEC).
In addition, we perform numerical calculations that confirm our result and we discuss its applicability to Hamiltonian dynamics.
Finally, we briefly explore the application of our result to the generation of approximate state designs and the construction of a novel quantum circuit compression scheme.

\paragraph{Setup.}
We consider a quantum circuit on $N$ qubits initialized in the $\ket{00\ldots0}$ state with alternating applications of deep random Clifford circuits and single T-gates~\footnote{
	The choice of T-gates on the first qubit is equivalent with any other choice $G_i = L_i T R_i$ for any Cliffords $L_i, R_i \in \mathcal{C}_N$ (where $\mathcal{C}_N$ denotes the set of Clifford unitarys on $N$ qubits) because they could be absorbed into the adjacent random Clifford circuits $C_i$ and $C_{i+1}$ respectively.
	In particular (because $\mathtt{SWAP}\in\mathcal{C}_2$) these circuits are equivalent to circuits with T-gates applied to any sequence of qubits.},
as exemplified in Fig.~\ref{fig:clifford-circuit}.
While this circuit leads to highly entangled states already after a single time step, the nonstabilizerness of the state will increase only slowly through the application of T-gates.
As the set of Cliffords together with T-gates is universal, these circuits may approach any state after $O(N)$ to $O(N^2)$ time steps~\cite{beverland2020lower, jiang2023lower, amy2019tcount}.

In this work we consider the tradeoff between nonstabilizerness and entanglement that arises if we commute the T-gates through the Clifford circuits onto the initial state.
This move can shift the nonstabilizerness onto the initial state, but generally at the expense of increasing its entanglement:
Consider the quantum circuit in Fig.~\ref{fig:clifford-circuit} up to time $t=1$, with $\ket{\psi(1)} = T C_1 \ket{00\ldots0}$.
Because $T=\e{-i(\pi/8)Z_1}$, we have $\ket{\psi(1)} = C_1 \tilde{T} \ket{00\ldots0}$, where $\tilde{T}=\e{-i(\pi/8)\tilde{Z}_1}$ with $\tilde{Z}_1 = C_1^\dagger Z_1 C_1$.
As $\tilde{Z}_1$ is a non-local random Pauli operator, $\tilde{T} \ket{00\ldots0}$ will generally be an entangled state.
We shall establish conditions under which one can disentangle such a state using a Clifford circuit $\tilde{C}_1$ such that $\ket{\psi(1)} = \tilde{C}_1 \ket{T00\ldots0}$, where $\ket{T}=\e{-i(\pi/8)Y} \ket{0}$.
If these conditions are fulfilled repeatedly for some $t\leq N$, we then may find Cliffords $\tilde{C}_t$ such that $\ket{\psi(t)}=  \tilde{C}_t \left( \ket{T}^{\otimes t}\ket{0}^{\otimes{N-t}} \right)$, as in Fig.~\ref{fig:clifford-circuit}.

\paragraph{Results.}
To establish the result in Fig.~\ref{fig:clifford-circuit}, we first seek conditions under which we can rewrite $\ket{\psi}=T C \left(\ket{T}^{\otimes k} \ket{0}^{\otimes N-k} \right)$ as $\ket{\psi}= \tilde{C} \left(\ket{T}^{\otimes k+1} \ket{0}^{\otimes N-k-1} \right)$.
To this end, we view the problem through the lens of QEC and consider a more general situation, shown in Fig.~\ref{fig:theorem}.
Here a gate $\e{i\phi P}$ with phase $\phi$ and $P\in\mathcal{P_N}$ (with $\mathcal{P}_N$ the $N$-qubit Pauli group) is applied to a state $\ket{\psi}$ of a stabilizer code wherein a (nonstabilizer) logical state $\ket{\varphi}$ is encoded through a Clifford unitary $C$.
This applies to Fig.~\ref{fig:clifford-circuit}
for $P = Z_1$ and $\phi=-\pi/8$.
The result in Fig.~\ref{fig:theorem} expresses the following:
\begin{theorem} \label{th:theorem}
	Let $\ket{\psi}= C \left( \ket{\varphi}\otimes\ket 0^{\otimes(N-k)}\right)$ be an $N$-qubit encoding of a $k$-qubit logical state $\ket{\varphi}$ through a Clifford unitary $C$, with $0 \leq k < N$.
	If Pauli operator $P\in\mathcal{P}_N$ is not a logical operator of the corresponding $[N,k]$ stabilizer code, then there exists a Clifford unitary $\tilde{C}$ such that (up to a global phase factor)
	\begin{equation} \label{eq:theorem}
		\e{i \phi P} \ket{\psi} \propto \tilde{C} \left(\ket{\varphi}\otimes\ket{x(\phi)}\otimes\ket{0}^{\otimes(N-k-1)} \right)\mathrm{,}
	\end{equation}
	where $\ket{x(\phi)}=\ket{0}$ or $\ket{x(\phi)} = \e{i \phi Y}\ket{0}$.
\end{theorem}

\begin{proof}
	If $P$ is not a logical operator then either $P\in \mathcal{S}$ or $PS_j=-S_jP$ for some $j>k$, where $S_i=CZ_iC^\dagger$ generate the code's stabilizer group $\mathcal{S}$ ($i\in\{k+1,\ldots, N\}$).
	If $P \in \mathcal{S}$, then $P \ket{\psi} = \ket{\psi}$ hence Eq.~\eqref{eq:theorem} holds with $\ket{x(\phi)}=\ket{0}$ and $\tilde{C}=C$.
	If $PS_j=-S_jP$ then $C^\dagger P C$ anticommutes with $Z_j$ hence $C^\dagger P C = L R$ where $R$ is either $X_j$ or $Y_j$ and  $L$ is a Pauli operator on all other qubits in $\ket 0^{\otimes(N-k)}$.
	We label qubits in $\ket 0^{\otimes(N-k)}$ such that $j=k+1$ and assume w.l.o.g.~that $R=X_{k+1}$.
	(If $R=Y_{k+1}$ we replace $C$ by $C^{\prime} = C \, \e{-i (\pi/4) Z_{k+1}}$ using $Z_{k+1}\ket 0^{\otimes(N-k)}=\ket 0^{\otimes(N-k)}$.)   
		Using \mbox{$V=e^{i(\pi/4) \, L Z_{k+1}}$} gives $L X_{k+1} = V \, Y_{k+1} \, V^\dagger$ and 
	\begin{equation}
		V^\dagger \left(\ket{\varphi}\otimes\ket{0}^{\otimes(N-k)} \right)= W^\dagger \left(\ket{\varphi}\otimes\ket{0}^{\otimes(N-k)} \right),
	\end{equation}
	with $W=e^{i(\pi/4) L}$.
	Hence
    \begin{subequations}
	\begin{align}
		\e{i \phi P} \ket{\psi} &= C e^{i\phi L X_{k+1}} \left( \ket{\varphi}\otimes\ket 0^{\otimes(N-k)}\right) \\
		&= C V \e{i\phi Y_{k+1}} W^\dagger \left( \ket{\varphi}\otimes\ket 0^{\otimes(N-k)}\right) \\
		&= C V W^\dagger \left( \ket{\varphi}\otimes \ket{x(\phi)} \otimes\ket{0}^{\otimes(N-k-1)}\right), \label{eq:construction}
	\end{align}
    \end{subequations}
	with $\ket{x(\phi)} = \e{i\phi Y}\ket{0}$.
	This proves Eq.~\eqref{eq:theorem} because $C$, $V$, and $W^\dagger$ are all Clifford operations.
\end{proof}
To use Theorem~\ref{th:theorem} to show the result in Fig.~\ref{fig:clifford-circuit}, we estimate the maximum $t^*$ such that $P=Z_1$ in the first $t^*$ T-gates is not a logical operator. For simplicity, we consider a process where in each such step a new logical $\ket{T}$ is encoded; this differs from Fig.~\ref{fig:clifford-circuit} for $Z_1\in\mathcal{S}$ and hence will slightly underestimate $t^*$~\footnote{The error is exponentially small in $N$: by our probabilistic argument below, $\text{Pr}(Z_1\in\mathcal{S})\leq 2^{N}/(4^N-1)$.}.  
Our estimate is probabilistic, using that deep random Clifford circuits generate random stabilizer codes. The probability that $P$ is not a logical for a random $[N,k]$ stabilizer code  (equivalently that a uniformly random $P$ is not a logical for a given $[N,k]$ stabilizer code) is $p_{k+1} = 1 - \frac{(4^k-1)\,2^{N-k}}{4^N -1}$ since there are $2^{N-k}$ representatives for each of the $4^k-1$ logical Pauli operators and there are $4^N -1$ non-trivial Pauli strings. 
Therefore, the probability $\text{Pr}(t^*)$ for $P$ in the first $t^*$ T-gates not to be a logical, but to be so in the $(t^*+1)^\text{th}$  is $\text{Pr}(t^*)=(1-p_{t^*+1})\prod_{k=1}^{t^*}p_{k}$.
This gives $\text{Pr}(N-j)\approx\left(\frac{1}{2};\frac{1}{2}\right)_{\infty}2^{-j}/\left(\frac{1}{2};\frac{1}{2}\right)_{j}$ for large $N$, where  $(a;q)_{n}=\prod_{k=0}^{n-1}(1-aq^{k})$.
Using this, $\langle \tau \rangle = \langle N-t^*\rangle\approx1.607$ with standard deviation $\sigma_\tau \approx1.6565$. 

Although we phrased our argument using T-gates, by Theorem~\ref{th:theorem} our result equally applies if in Fig.~\ref{fig:clifford-circuit}, instead of a T-gates, we use $G_j=\e{i\phi_j P_j}$, with any $P_j\in\mathcal{P}_N$, as the $j^\text{th}$ non-Clifford gate.
The (explicit) ansatz  $\ket{\psi(t)} =  \tilde{C}_t \left( \ket{x(\phi_1)\ldots x(\phi_t)}\ket{0}^{\otimes{N-t}} \right)$ we found, however applies only if $P_j$ are not logical operators and $t<t^*$.
To go beyond this case, we represent  $\ket{\psi(t)}$ as a Clifford augmented matrix product state (CAMPS)~\cite{masotllima2024stabilizer, qian2024augmenting, huang2024nonstabilizerness}: as an MPS acted upon by a Clifford circuit $C$, i.e. $\ket{\psi} = C \MPS$.
To establish the general features of this CAMPS, we consider $\ket{\psi(t)}= G_t C_t \ldots G_1 C_1 \left( \ket{0}^{\otimes N} \right)$ and in each intermediate time step $s \leq t$ apply Theorem~\ref{th:theorem}. This yields states of the form
\begin{equation}\label{eq:compression-form}
\ket{\psi(s)}= \tilde{C}_s \left(\textstyle{\ket{\mathtt{MPS}^{(k)}}}\ket{0}^{\otimes N-k} \right),
\end{equation}
where $k\leq s$, and $\ket{\mathtt{MPS}^{(k)}}$ is an MPS on only the first $k$ qubits.
This is because for each $s$ either $P_{s+1}$ is a logical  of the $[N,k]$ code given by $C:= C_{s+1} \tilde{C}_s$, or it is not.
If it is, then $C^\dagger P_{s+1} C$ acts trivially on the last $N-k$ qubits and  $\ket{\psi(s+1)}= \tilde{C}_{s+1} \left(\ket{\mathtt{MPS}^{\prime(k)}}\ket{0}^{\otimes N-k} \right)$, with $\tilde{C}_{s+1} = C$ and  $\ket{\mathtt{MPS}^{\prime(k)}} = \e{i\phi_s C^\dagger P C} \ket{\mathtt{MPS}^{(k)}}$ of doubled bond dimension.
Otherwise, 
we can use $\tilde{C}$ from  Theorem~\ref{th:theorem} to get $\ket{\psi(s+1)}= \tilde{C}_{s+1} \left(\ket{\mathtt{MPS}^{(k+1)}}\ket{0}^{\otimes N-k-1} \right)$, with $\tilde{C}_{s+1} = \tilde{C}$ and $\ket{\mathtt{MPS}^{(k+1)}} = \ket{\mathtt{MPS}^{(k)}}\ket{x(\phi_s)}$.
Hence the final state $\ket{\psi(t)}$ will be as in Eq.~\eqref{eq:compression-form}, with $k \leq t$ and MPS bond dimension at most $2^{t-k}$.
While $k$ depends on the Clifford operations $C_1, \ldots, C_t$, we expect $t-k = O(1)$ for generic deep (i.e., global and random) Clifford circuits and $t<t^*$. For $t>t^*$, $k$ reaches $N$ and we expect the bond dimension to rapidly increase with $t$. 

To verify our predictions, we now study the setup in Fig.~\ref{fig:clifford-circuit} numerically.
For this, we employ a CAMPS representation of the state and apply a variant of the so-called ``entanglement cooling'' algorithm first introduced in Refs.~\cite{chamon2014emergent,shaffer2014irreversibility}, and recently reformulated in Refs.~\cite{true2022transitions, mello2024hybrid, qian2024augmenting, mello2024clifford, qian2024clifford, nakhl2024stabilizer, liu2024classical}.
When applying a Clifford unitary to a CAMPS $\ket{\psi} = C \MPS$ we update $C$ using the stabilizer formalism. 
We apply a T-gate by commuting it with $C$, obtaining \mbox{$\tilde{T} = \alpha \id + \beta \tilde{Z}$}, where  $\alpha=\cos(\pi/8)$, $\beta=-i\sin(\pi/8)$, and $\tilde{Z}= C^\dagger Z C$.
Then we write $\tilde{T}$ as a matrix product operator (MPO) with bond dimension two and contract it with the MPS to get a new CAMPS
\begin{equation}
\ket{\tilde{\psi}} = T \ket{\psi} = T C \MPS = C \tilde{T} \MPS = C \ket{\widetilde{\mathtt{MPS}}} \mathrm{.}
\end{equation}
This, however, generically doubles the MPS bond dimension.
Therefore, after each such step we apply a Clifford unitary $Q$ such that the entanglement of $Q \MPS$  is minimized.
Then we rewrite $\ket{\psi} = C \MPS = C Q^\dagger Q \MPS = C^\prime \MPS[^\prime]$ where both $C^\prime = C Q^\dagger$ and $\MPS[^\prime] = Q \MPS$ can be computed in $\mathrm{poly}(N)$ time using the stabilizer formalism and tensor network contractions respectively, provided $Q$ can be decomposed into a sequence of shallow local quantum circuits that each do not increase the bond dimension of the MPS.

If the conditions of Theorem~\ref{th:theorem} hold, we could use its $\tilde{C}$ and set $Q=\tilde{C}^\dagger \, C$. 
However, for more general applicability, we optimize $Q$ numerically.
For optimizing $Q$, we found a ``greedy sweep'' search to perform best.
For this, we start by applying trial two-qubit Clifford unitaries to the first two qubits and for each record the entanglement entropy (EE) between the first qubit and the rest.
Using the Clifford $V_1$ minimizing EE, we update as $\MPS[^\prime] = V_1 \MPS$.
Then we apply trial Cliffords to qubit two and three, obtaining $V_2$,  update the MPS, and so on.
Upon reaching the last two qubits we sweep back until reaching the first two qubits.
This is then repeated until no further improvement is achieved.
The resulting Clifford unitary is $Q =  \ldots V_2 V_1$.
For this search, we use only such two-qubit Cliffords that may reduce EE and cannot be transformed into one another via local operations: the quotient $\tilde{\mathcal{C}}_2 = \mathcal{C}_2 / (\mathcal{C}_1 \otimes \mathcal{C}_1)$, with $\mathcal{C}_n$ the $n$-qubit Clifford group.
As $|\mathcal{C}_1| = 24$ and $|\mathcal{C}_2| = 11\,520$ there are only $|\tilde{\mathcal{C}}_2| = 20$ Cliffords to be checked at each step.
We find that to minimize EE typically only $O(1)$ sweeps are needed and thus the computation time to find $Q$ in our systems scales linearly with $N$.

We now show numerical results using EE and the stabilizer R\'{e}nyi entropy (SRE) as measures for entanglement and nonstabilizerness, respectively.
The EE of a pure $N$-qubit state $\ket{\psi}$ for subsystem $L$ is
\begin{equation}
	\mathcal{E}_i(\ket{\psi}) = - \mathrm{tr}\left[ \rho_L \log_2(\rho_L) \right]\mathrm{,}
\end{equation}
where $\rho_L = \mathrm{tr}_R\left[\ketbra{\psi}{\psi}\right]$ is the reduced density matrix for qubits $L=\{1, \ldots, i\}$ with $\mathrm{tr}_R$ the partial trace over sites $R = \{i+1, \ldots, N\}$.
For all results below we will consider the maximum \mbox{$\mathcal{E}(\ket{\psi}) = \max \left[ \{ \mathcal{E}_i(\ket{\psi}) \}_{i\in 1\ldots N} \right]$}.
The SRE~\cite{leone2022stabilizerrenyientropy, lami2023nonstabilizerness} quantifies the spread of a state $|\psi\rangle$ in the basis of Pauli string operators. It is given by
\begin{equation} \label{eq:sre-def}
	\mathcal{M}(|\psi\rangle) = - \log_2 \left( \sum_{P\in \mathcal{P}_N} \Xi_P^2(|\psi\rangle) \right) - N \mathrm{,}
\end{equation}
where $\Xi_P(|\psi\rangle) = 2^{-N} \langle\psi| P |\psi\rangle^2$ is the probability distribution over all Pauli strings $P \in \mathcal{P}_N$ with +1 phases.

\begin{figure}
	\phantomsubfloat{\label{fig:clifford-EE-vs-time}}%
	\phantomsubfloat{\label{fig:clifford-SRE-vs-time}}%
	\centering
	\includegraphics[width=1\linewidth]{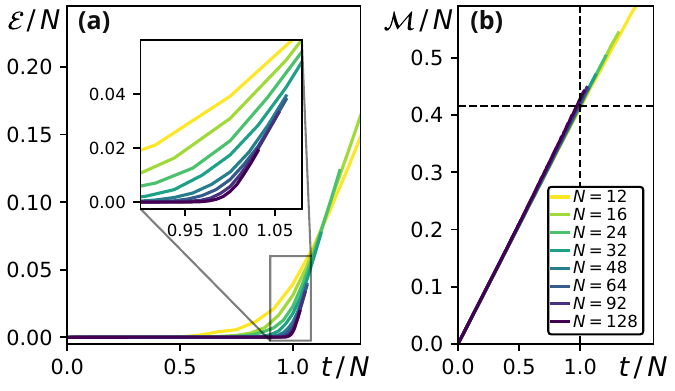} 
	\caption{\label{fig:clifford-XE-vs-time}%
		The evolution of the maximal entanglement entropy density of the MPS ($\mathcal{E}/N$) of the CAMPS  [panel (a)] and the stabilizer R\'{e}nyi entropy density ($\mathcal{M}/N$) for the Clifford + T dynamics [panel (b)] for different system sizes $N$, averaged over 256 random Clifford sequences.
		Panel~(a) shows that the states generated by the Clifford + T circuits can be almost completely disentangled for the first $N$ time steps.
		The horizontal dashed line in panel~(b) marks the SRE density of $\ket{T}^{\otimes N}$. 
	}
\end{figure}

Figure~\ref{fig:clifford-XE-vs-time} shows numerical results averaged over 256 random Clifford sequences $C_1, \ldots, C_t$ for different system sizes $N$.
We construct $C_j$ by applying $2N^2$ two-qubit Clifford gates on random (non-local) pairs of qubits; each gate is sampled independently and uniformly from $\mathcal{C}_2$.
We find that we can disentangle the state completely for $t\lesssim N$~\footnote{Reference~\cite{lami2024quantum} mentions a similar numerical observation in its appendix.}. 
For $t \gtrsim N$, the MPS bond dimension increases rapidly with $t$, aligning with our expectation above.
The transition from a completely disentangled to an entangled MPS becomes sharper as $N$ increases.
For large $N$, we find that we can \emph{fully} disentangle for up to $t^* = N - \tau$ time steps, with average $\langle \tau \rangle = 1.61\pm0.09$ and standard deviation $\sigma_\tau = 1.60\pm0.12$.
This is in perfect agreement with our above analytical results.

\begin{figure}
	\centering
	\includegraphics[width=1\linewidth]{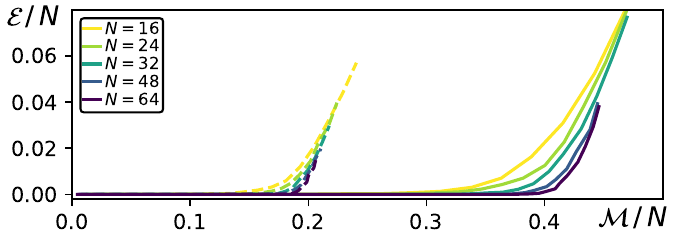} 
	\caption{\label{fig:clifford-EE-vs-SRE}%
		The maximal EE density of the MPS of the CAMPS ansatz versus the SRE density for the Clifford + T-gate circuit (solid lines) and the Clifford + $\sqrt{\mathrm{T}}$-gate circuit (dashed lines).
		These results follow from \mbox{$\mathcal{M}(\sqrt{T}\ket{+})\approx 0.2075$} being the half of \mbox{$\mathcal{M}(T\ket{+}) \approx 0.4150$} and that in both cases the MPS of the CAMPS ansatz can be completely disentangled for approximately $N$ time steps but not further.
	}
\end{figure}

Recalling that our results apply also to $G_j=\e{i\phi_j P_j}$ as the $j^\text{th}$ non-Clifford gate, we now study the system with   $\sqrt{\mathrm{T}}$-gates as $G_j$.
We find  the same $\langle t^* \rangle$ as with T-gates.
As $\mathcal{M}(\sqrt{T}\ket{+}) =  \frac{1}{2} \mathcal{M}(T\ket{+}) \approx \frac{1}{2} \times 0.4150 = 0.2075$, the maximal SRE that we can simulate using CAMPS, while still maintaining an MPS of bond dimension one, has been cut in half, as is shown in Fig.~\ref{fig:clifford-EE-vs-SRE}.

\paragraph{Consequences for Hamiltonian dynamics.}
To relate our results to Hamiltonian dynamics, we consider a generic Hamiltonian $H = \sum_{j=1}^M \omega_j P_j$, with $M>N$ Pauli strings $P_j$ and energies $\omega_j$.
For the evolution for a small time step $\dt$ under a Suzuki-Trotter decomposition~\cite{suzuki1992general} $\e{-iH\dt} \approx \prod_{j=1}^{M} \e{-i\omega_j\dt P_j}$, we already need $M$ non-Clifford gates.
This suggests that Hamiltonian dynamics will generically not admit a completely disentangled CAMPS representation, not even at early times.
In the End Matter, we present numerical results confirming this for a non-integrable one-dimensional Ising chain.

\paragraph{Applications.}
Our results, firstly, suggest a novel and efficiently implementable representation of approximate state designs.
For this, we use that random global Clifford circuits alternated with a uniformly sampled ensemble of $\{K, K^\dagger, \id\}$ gates (where $K$ is any non-Clifford single qubit gate) generate $\epsilon$-approximate unitary $k$-designs when using $t=O[k^4 \log^2(k) \log(1/\epsilon)]$ layers~\cite{haferkamp2023efficient}.
Furthermore, a unitary $k$-design acting on any state---including $\ket{00\ldots0}$---yields a state $k$-design~\cite{mele2024introduction}.
Choosing \mbox{$K=\e{i \alpha Z_1}$} with $\alpha\neq j\pi/4$ $(j\in \mathbb{Z})$, we can hence use our above results to generate $\epsilon$-approximate state $k$-designs by applying random global Cliffords to the ensemble  $\ket{x(\phi_1)\,x(\phi_2)\,\ldots\,x(\phi_t)}\ket{0}^{\otimes N-t}$ with $\phi_j$ sampled uniformly from $\{\alpha, -\alpha, 0 \}$%
~\footnote{Contributions to the state design that cannot be transformed into this form are suppressed exponentially in $N-t$.}%
~\footnote{By Remark 1 in Ref.~\cite{haferkamp2023efficient}, if $K^2$ is non-Clifford, then $\phi_j$ may be sampled from just $\{\alpha, -\alpha \}$}.
It was recently shown that random global Cliffords applied to random matrix product states with bond dimension $\chi$ generate $\epsilon$-approximate state $4$-designs with \mbox{$\epsilon = O(1/\chi^2)$~\cite{lami2024quantum}}.
Complementarily, the above shows that random global Cliffords applied to an ensemble of magic product states can generate $\epsilon$-approximate $k$-designs with \mbox{$N > t = O[k^4 \log^2(k) \log(1/\epsilon)]$}.

Further applications include the classical computation of Pauli expectation values and  a quantum circuit compression scheme for sampling tasks, both for the setup in Fig.~\ref{fig:clifford-circuit}  with $t<N$ and any non-Clifford phase gates $G_j = \e{i\phi_j P_j}$, not just $T$.
Using $\ket{\psi(t)}$ in the CAMPS form of Eq.~\eqref{eq:compression-form}, we can calculate the expectation of any $P\in\mathcal{P}_N$ in $\mathrm{poly}(N)$ time.
For this we use that $\tilde{P} = \tilde{C}_t^\dagger P \tilde{C}_t\in\mathcal{P_N}$ and extract the expectation by contraction with the MPS, i.e.  $\bra{\psi}P\ket{\psi}= \left(\bra{\mathtt{MPS}^{(k)}} \bra{0}^{\otimes N-k}\right)\tilde{P}\left(\ket{\mathtt{MPS}^{(k)}} \ket{0}^{\otimes N-k} \right)$.

While computing Pauli expectations is classically efficient, sampling the distribution of  projective Pauli measurements is harder.
However, as we show, for the setup in Fig.~\ref{fig:clifford-circuit} with $t<N-O(1)$ we need only a $k\leq t$ qubit quantum computer. Suppose we want to sample output bit strings for projective measurements of $Z_1, \ldots, Z_N$ on $\ket{\psi(t)}$.
Using the form of $\ket{\psi(t)}$ from Eq.~\eqref{eq:compression-form}, this is the same as sampling $\tilde{Z}_1, \ldots, \tilde{Z}_N$ from $\MPS[^{(k)}]\ket{0}^{\otimes N-k}$, where $\tilde{Z}_j = \tilde{C}_t^\dagger Z_j \tilde{C}_t\in\mathcal{P}_N$.
For any $\tilde{Z}_j$ that anticommutes with $Z_{j>k}$, the measurement outcome is $+1$ or $-1$ with equal probability, and the post-measurement state is obtained via a Clifford unitary~\cite{bravyi2016trading}.
Commuting these out of the circuit, we are left with  at most $k$ mutually commuting Pauli measurements, which we can then restrict to $\ket{\mathtt{MPS}^{(k)}}$ (since they have to feature $\id$ or $Z_j$ on qubits $j>k$ to be retained)~\cite{bravyi2016trading}.
Hence we restricted the quantum part of the problem to $n\leq k$ qubits.

If, for example, we take $t = N/2$ and $t-k = O(1)$, the original task is reduced to non-local sampling on at most $N/2$ qubits from a state created by a shallow circuit of depth $O(1)$.
This approach is akin to Pauli based computation (PBC)~\cite{bravyi2016trading}.
However, while PBC is restricted to T-gates, our scheme allows for any non-Clifford $G_j = \e{i\phi_j P_j}$, as long as the Clifford blocks are global random, i.e., deep random Clifford circuits.

\paragraph{Discussion and conclusion.}
We have established conditions under which one can transform multiple layers of Clifford circuits interspersed with general non-Clifford phase gates into a single Clifford circuit applied to a magic product state.
We formulated these conditions in terms of QEC and found that for random Clifford circuits they are typically fulfilled for up to $\langle t^* \rangle \approx N-1.607$ interspersed non-Clifford phase gates.
Building on these results we proposed a protocol to generate state $k$-designs and a circuit compression scheme.
Our results extend beyond Clifford + T circuits to include arbitrary-angle non-Clifford rotations, which would require deeper circuit architectures and a greater number of non-Clifford gates if instead synthesized from T-gates.
While we demonstrate that for the quantum circuits considered here Pauli string expectation values can be computed classically in polynomial time, sampling from $O(N)$ output qubits seems to necessitate exponential classical resources.
This is because the bond dimension of the MPS in the CAMPS representation generically doubles with each output qubit projection.
However, akin to Instantaneous Quantum Polytime (IQP) circuits~\cite{shepherd2009temporally,bremner2011classical,bremner2016averagecase}, sampling from $O[\log(N)]$ qubits is still classically efficient.
Future work could explore the worst case and average case performance of the proposed method using the CAMPS ansatz, and study its connection to the onset of quantum chaos~\cite{chamon2014emergent, shaffer2014irreversibility, true2022transitions, leone2024learning, haug2024probing}, and to the classical simulation complexity of quantum circuits with shallow magic depth~\cite{zhang2025classical}. 

\vspace{5mm}
\paragraph{Acknowledgements.}
We thank Alioscia Hamma, Tobias Haug, Xhek Turkeshi, Piotr Sierant, and especially Marcello Dalmonte, and Mario Collura for helpful discussions.
We thank Poetri Tarabunga for pointing out that only $20$ Clifford gates are necessary to reduce the entanglement in CAMPS.
G.\,E.\,F., E.\,T., and R.\,F. acknowledge support from  ERC under grant agreement n.101053159 (RAVE).
B.\,B. acknowledges support through EPSRC grant EP/V062654/1.
R.\,F. was partly supported by the PNRR MUR project PE0000023-NQSTI.

\subsection{End Matter}
In the main text of this work we found that CAMPS performs particularly well for Clifford circuits interspersed with general non-Clifford phase gates. 
We now contrast this result with CAMPS applied to the Hamiltonian dynamics of the quantum Ising model, choosing longitudinal fields
that break integrability. 
We find that the transient state of such an evolution can only be partly disentangled with Clifford circuits at very early times.
This is consistent with the insights gained from the alternating Clifford and phase gate circuits and suggests that generic Hamiltonian dynamics does not profit from the CAMPS ansatz.
Finally, we briefly study the use of matchgate circuits, instead of Cliffords, alongside tensor networks to reduce the entanglement but also the nonstabilizerness of Hamiltonian evolution of many-body quantum systems near free-fermion integrability.

\begin{figure}
	\phantomsubfloat{\label{fig:hamiltonian-EE-vs-time-N}}%
	\phantomsubfloat{\label{fig:hamiltonian-SRE-vs-time-N}}%
	\centering
	\includegraphics[width=1\linewidth]{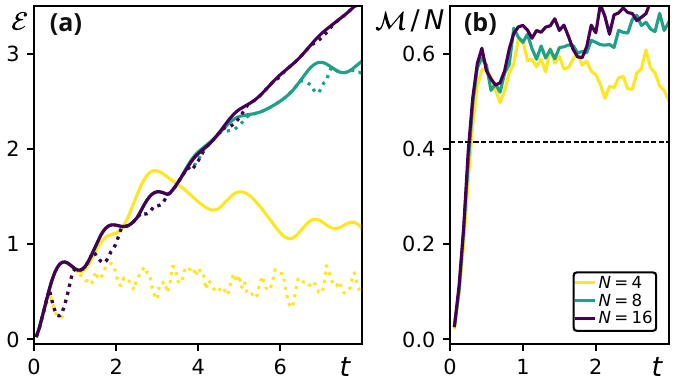} 
	\caption{\label{fig:hamiltonian-XE-vs-time-N}%
		The evolution of the maximal EE (a) and the SRE density (b) generated by the Ising Hamiltonian of Eq.\eqref{eq:hamiltonian-ising} for $h_x=0.3$ and different system sizes $N$.
		The solid lines in panel~(a) show the EE of the fully evolved state while the dotted lines show the EE of the MPS of the CAMPS ansatz.
		We notice that the CAMPS ansatz only partially disentangled the state at certain very early times or for very small system sizes.
		Also, we can see in panel (b) that the SRE density quickly grows and saturates above the SRE density of the $\ket{T}^{\otimes N}$ state, indicated with the dashed horizontal line.
	}
\end{figure}

\paragraph{Hamiltonian dynamics.}
We consider the Hamiltonian dynamics of a one-dimensional Ising chain with $N$ sites as a prototypical model with a Hamiltonian given by
\begin{equation}\label{eq:hamiltonian-ising}
	H = J \sum_{i=1}^{N-1} X_{i} X_{i+1} + h_{x} \sum_{i=1}^N X_{i} + h_{z} \sum_{i=1}^N Z_{i}.
\end{equation}
We set $\hbar=1$ and express all frequencies in terms of some arbitrary reference, such that the interaction strength is $J=1.0$, and the magnetic field with X and Z components is $h_x \geq 0.0$ and $h_z = 0.5$.
To evolve the initial $\ket{y+}^{\otimes N} \propto (\ket{0} +i\ket{1})^{\otimes N}$ state using the CAMPS ansatz we use an adapted version of time-dependent variational principle~(TDVP)~\cite{haegeman2011timedependent} as it has been introduced in Refs.~\cite{mello2024clifford, qian2024clifford}.
For this, we transform the Hamiltonian $H$ with the Clifford part $C(t)$ of the CAMPS $\ket{\psi(t)} = C \MPS$ to yield $\tilde{H} = C^\dagger H C$.
This can be done in $\mathrm{poly}(N)$ time because $H$ is a linear combination of $3N-1$ Pauli strings, each of which can be transformed efficiently using the stabilizer formalism.
This transformed Hamiltonian can then be written as an MPO with a bond dimension of at most $3N-1$ and then the canonical TDVP algorithm can be used to evolve the MPS with $\exp(-i \tilde{H} \dt)$, for some short time step $\dt$, such that
\begin{eqnarray}
	\ket{\psi(t+\dt)} &=& e^{-i H \dt} \ket{\psi(t)} = e^{-i H \dt} C \MPS \\
	&=& C e^{-i \tilde{H} \dt} \MPS = C \MPS[^\prime] \mathrm{.}
\end{eqnarray}
Again, after such an evolution step the entanglement in the MPS may increase and we perform a search for a suitable $Q \in \mathcal{C}_N$ such that $Q \MPS[^\prime]$ has a lower entanglement and then update the representation of the state accordingly.

\begin{figure}
	\phantomsubfloat{\label{fig:hamiltonian-EE-vs-time-hx}}%
	\phantomsubfloat{\label{fig:hamiltonian-SRE-vs-time-hx}}%
	\centering
	\includegraphics[width=1.0\linewidth]{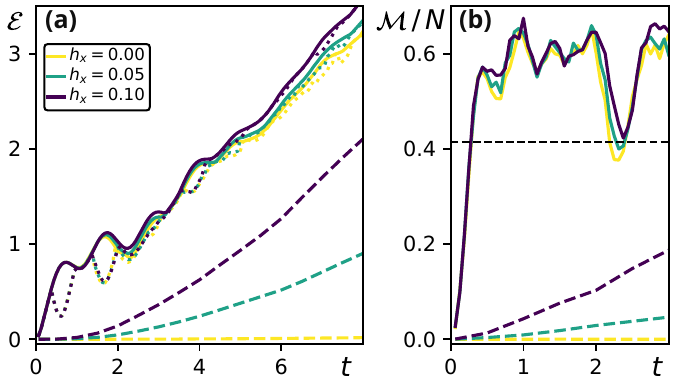} 
	\caption{\label{fig:hamiltonian-XE-vs-time-hx}%
		The evolution of the maximal EEW and the SRE density generated by the Ising Hamiltonian of Eq.\eqref{eq:hamiltonian-ising} for different values of $h_x$ and $N=16$.
		The solid lines in panels~(a)~and~(b) show the entropy of the evolved state, while the dotted and dashed lines show the entropy of the MPS of the CAMPS ansatz and of the state back-propagated with the matchgate circuit $\bar{U}^\dagger(t)$, respectively.
		The horizontal dashed line in panel~(b) marks the SRE density of the $\ket{T}^{\otimes N}$ state. 
	}
\end{figure}

Figure~\ref{fig:hamiltonian-EE-vs-time-N} shows the result of such an evolution for short times and three different system sizes $N$.
We only find Clifford unitaries that partially disentangle the state at certain very early times or for very small system sizes.
Any reduction of entanglement seems to disappear for larger systems and later times.
Furthermore, we notice that the state cannot be disentangled completely immediately after the first time step, which is consistent with the picture gained in the Clifford + T circuit above because the preparation of the state at time $t=\dt$ already would require $3N-1$ gates of the form $\e{i\phi_j P_j}$ where $P_j$ are the Pauli strings appearing in Eq.~\eqref{eq:hamiltonian-ising}.

\paragraph{Matchgate disentangling.}
Alternatively to disentangling with Clifford circuits, we now consider the use of matchgates \cite{valiant2001quantumcomputers, wu2025disentangling}; these include the evolution with Eq.~\eqref{eq:hamiltonian-ising} for the case of $h_x = 0$.
As a crude estimate for the utility of such a method, we consider disentangling $\ket{\psi(t)}$ as $\bar{U}^\dagger(t) \ket{\psi(t)} = \exp\left( +i \bar{H} t \right) \ket{\psi(t)}$, where $\bar{H}$ is $H$ from Eq.~\eqref{eq:hamiltonian-ising} with $h_x=0$ such that $\bar{U}^\dagger(t)$ is a matchgate circuit.  
In Fig.~\ref{fig:hamiltonian-EE-vs-time-hx} we plot the entanglement in $\ket{\psi(t)}$ as solid lines for different $h_x$ with system size $N=16$.
The dotted lines, again show the entanglement of the MPS part of a CAMPS ansatz after a search for the best entangling Cliffords, and the dashed lines show the entanglement entropy of $\bar{U}^\dagger(t) \ket{\psi(t)}$.
While we used a specific matchgate circuit $\bar{U}^\dagger(t)$ there may exist other matchgate circuits $M$ that could disentangle $\ket{\psi(t)}$ more efficiently, in a similar spirit to CAMPS, but optimizing over matchgate circuits $M$ instead of Cliffords.
With matchgate disentanglers, one may also consider tracking a fermionic form of magic~\cite{hebenstreit2019allpurefermionic} instead of magic relative to stabilizer states.
We leave exploring this to future work.


%


\begin{thebibliography}{67}%
	\makeatletter
	\providecommand \@ifxundefined [1]{%
		\@ifx{#1\undefined}
	}%
	\providecommand \@ifnum [1]{%
		\ifnum #1\expandafter \@firstoftwo
		\else \expandafter \@secondoftwo
		\fi
	}%
	\providecommand \@ifx [1]{%
		\ifx #1\expandafter \@firstoftwo
		\else \expandafter \@secondoftwo
		\fi
	}%
	\providecommand \natexlab [1]{#1}%
	\providecommand \enquote  [1]{``#1''}%
	\providecommand \bibnamefont  [1]{#1}%
	\providecommand \bibfnamefont [1]{#1}%
	\providecommand \citenamefont [1]{#1}%
	\providecommand \href@noop [0]{\@secondoftwo}%
	\providecommand \href [0]{\begingroup \@sanitize@url \@href}%
	\providecommand \@href[1]{\@@startlink{#1}\@@href}%
	\providecommand \@@href[1]{\endgroup#1\@@endlink}%
	\providecommand \@sanitize@url [0]{\catcode `\\12\catcode `\$12\catcode
		`\&12\catcode `\#12\catcode `\^12\catcode `\_12\catcode `\%12\relax}%
	\providecommand \@@startlink[1]{}%
	\providecommand \@@endlink[0]{}%
	\providecommand \url  [0]{\begingroup\@sanitize@url \@url }%
	\providecommand \@url [1]{\endgroup\@href {#1}{\urlprefix }}%
	\providecommand \urlprefix  [0]{URL }%
	\providecommand \Eprint [0]{\href }%
	\providecommand \doibase [0]{https://doi.org/}%
	\providecommand \selectlanguage [0]{\@gobble}%
	\providecommand \bibinfo  [0]{\@secondoftwo}%
	\providecommand \bibfield  [0]{\@secondoftwo}%
	\providecommand \translation [1]{[#1]}%
	\providecommand \BibitemOpen [0]{}%
	\providecommand \bibitemStop [0]{}%
	\providecommand \bibitemNoStop [0]{.\EOS\space}%
	\providecommand \EOS [0]{\spacefactor3000\relax}%
	\providecommand \BibitemShut  [1]{\csname bibitem#1\endcsname}%
	\let\auto@bib@innerbib\@empty
	\bibitem [{\citenamefont {Schollwöck}(2011)}]{schollwock2011thedensitymatrix}%
	\BibitemOpen
	\bibfield  {author} {\bibinfo {author} {\bibfnamefont {U.}~\bibnamefont
			{Schollwöck}},\ }\bibfield  {title} {\bibinfo {title} {The density-matrix
			renormalization group in the age of matrix product states},\ }\href
	{https://doi.org/10.1016/j.aop.2010.09.012} {\bibfield  {journal} {\bibinfo
			{journal} {Ann. Phys. (N. Y.)}\ }\textbf {\bibinfo {volume} {326}},\ \bibinfo
		{pages} {96} (\bibinfo {year} {2011})}\BibitemShut {NoStop}%
	\bibitem [{\citenamefont {Orús}(2019)}]{orus2019tensor}%
	\BibitemOpen
	\bibfield  {author} {\bibinfo {author} {\bibfnamefont {R.}~\bibnamefont
			{Orús}},\ }\bibfield  {title} {\bibinfo {title} {Tensor networks for complex
			quantum systems},\ }\href {https://doi.org/10.1038/s42254-019-0086-7}
	{\bibfield  {journal} {\bibinfo  {journal} {Nat. Rev. Phys.}\ }\textbf
		{\bibinfo {volume} {1}},\ \bibinfo {pages} {538} (\bibinfo {year}
		{2019})}\BibitemShut {NoStop}%
	\bibitem [{\citenamefont {Ran}\ \emph {et~al.}(2020)\citenamefont {Ran},
		\citenamefont {Tirrito}, \citenamefont {Peng}, \citenamefont {Chen},
		\citenamefont {Tagliacozzo}, \citenamefont {Su},\ and\ \citenamefont
		{Lewenstein}}]{ran2020tensor}%
	\BibitemOpen
	\bibfield  {author} {\bibinfo {author} {\bibfnamefont {S.-J.}\ \bibnamefont
			{Ran}}, \bibinfo {author} {\bibfnamefont {E.}~\bibnamefont {Tirrito}},
		\bibinfo {author} {\bibfnamefont {C.}~\bibnamefont {Peng}}, \bibinfo {author}
		{\bibfnamefont {X.}~\bibnamefont {Chen}}, \bibinfo {author} {\bibfnamefont
			{L.}~\bibnamefont {Tagliacozzo}}, \bibinfo {author} {\bibfnamefont
			{G.}~\bibnamefont {Su}},\ and\ \bibinfo {author} {\bibfnamefont
			{M.}~\bibnamefont {Lewenstein}},\ }\href
	{https://doi.org/10.1007/978-3-030-34489-4} {\emph {\bibinfo {title} {Tensor
				{Network} {Contractions}: {Methods} and {Applications} to {Quantum}
				{Many}-{Body} {Systems}}}},\ \bibinfo {series} {Lecture {Notes} in
		{Physics}}, Vol.\ \bibinfo {volume} {964}\ (\bibinfo  {publisher} {Springer
		International Publishing},\ \bibinfo {address} {Cham},\ \bibinfo {year}
	{2020})\BibitemShut {NoStop}%
	\bibitem [{\citenamefont {Vidal}(2003)}]{vidal2002efficient}%
	\BibitemOpen
	\bibfield  {author} {\bibinfo {author} {\bibfnamefont {G.}~\bibnamefont
			{Vidal}},\ }\bibfield  {title} {\bibinfo {title} {Efficient classical
			simulation of slightly entangled quantum computations},\ }\href
	{https://doi.org/10.1103/PhysRevLett.91.147902} {\bibfield  {journal}
		{\bibinfo  {journal} {Phys. Rev. Lett.}\ }\textbf {\bibinfo {volume} {91}},\
		\bibinfo {pages} {147902} (\bibinfo {year} {2003})}\BibitemShut {NoStop}%
	\bibitem [{\citenamefont {Vidal}(2004)}]{vidal2004efficient}%
	\BibitemOpen
	\bibfield  {author} {\bibinfo {author} {\bibfnamefont {G.}~\bibnamefont
			{Vidal}},\ }\bibfield  {title} {\bibinfo {title} {Efficient simulation of
			one-dimensional quantum many-body systems},\ }\href
	{https://doi.org/10.1103/PhysRevLett.93.040502} {\bibfield  {journal}
		{\bibinfo  {journal} {Phys. Rev. Lett.}\ }\textbf {\bibinfo {volume} {93}},\
		\bibinfo {pages} {040502} (\bibinfo {year} {2004})}\BibitemShut {NoStop}%
	\bibitem [{\citenamefont {Eisert}\ \emph {et~al.}(2010)\citenamefont {Eisert},
		\citenamefont {Cramer},\ and\ \citenamefont {Plenio}}]{eisert2010colloquium}%
	\BibitemOpen
	\bibfield  {author} {\bibinfo {author} {\bibfnamefont {J.}~\bibnamefont
			{Eisert}}, \bibinfo {author} {\bibfnamefont {M.}~\bibnamefont {Cramer}},\
		and\ \bibinfo {author} {\bibfnamefont {M.~B.}\ \bibnamefont {Plenio}},\
	}\bibfield  {title} {\bibinfo {title} {Colloquium: Area laws for the
			entanglement entropy},\ }\href {https://doi.org/10.1103/RevModPhys.82.277}
	{\bibfield  {journal} {\bibinfo  {journal} {Rev. Mod. Phys.}\ }\textbf
		{\bibinfo {volume} {82}},\ \bibinfo {pages} {277} (\bibinfo {year}
		{2010})}\BibitemShut {NoStop}%
	\bibitem [{\citenamefont {Gottesman}(2004)}]{gottesman2004stabilizer}%
	\BibitemOpen
	\bibfield  {author} {\bibinfo {author} {\bibfnamefont {D.~E.}\ \bibnamefont
			{Gottesman}},\ }\emph {\bibinfo {title} {Stabilizer {Codes} and {Quantum}
			{Error} {Correction}}},\ \href {https://doi.org/10.7907/RZR7-DT72} {Ph.D.
		thesis},\ \bibinfo  {school} {California Institute of Technology} (\bibinfo
	{year} {2004})\BibitemShut {NoStop}%
	\bibitem [{\citenamefont
		{Gottesman}(1998)}]{gottesman1998theoryoffaulttolerant}%
	\BibitemOpen
	\bibfield  {author} {\bibinfo {author} {\bibfnamefont {D.}~\bibnamefont
			{Gottesman}},\ }\bibfield  {title} {\bibinfo {title} {Theory of
			fault-tolerant quantum computation},\ }\href
	{https://doi.org/10.1103/PhysRevA.57.127} {\bibfield  {journal} {\bibinfo
			{journal} {Phys. Rev. A}\ }\textbf {\bibinfo {volume} {57}},\ \bibinfo
		{pages} {127} (\bibinfo {year} {1998})}\BibitemShut {NoStop}%
	\bibitem [{\citenamefont {Gottesman}(1999)}]{gottesmann1998faulttolerant}%
	\BibitemOpen
	\bibfield  {author} {\bibinfo {author} {\bibfnamefont {D.}~\bibnamefont
			{Gottesman}},\ }\bibfield  {title} {\bibinfo {title} {Fault-tolerant quantum
			computation with higher-dimensional systems},\ }\href
	{https://doi.org/10.1016/S0960-0779(98)00218-5} {\bibfield  {journal}
		{\bibinfo  {journal} {Chaos Solit. Fractals}\ }\textbf {\bibinfo {volume}
			{10}},\ \bibinfo {pages} {1749} (\bibinfo {year} {1999})}\BibitemShut
	{NoStop}%
	\bibitem [{\citenamefont {Aaronson}\ and\ \citenamefont
		{Gottesman}(2004)}]{aaronson2004improvedsimulation}%
	\BibitemOpen
	\bibfield  {author} {\bibinfo {author} {\bibfnamefont {S.}~\bibnamefont
			{Aaronson}}\ and\ \bibinfo {author} {\bibfnamefont {D.}~\bibnamefont
			{Gottesman}},\ }\bibfield  {title} {\bibinfo {title} {Improved simulation of
			stabilizer circuits},\ }\href {https://doi.org/10.1103/PhysRevA.70.052328}
	{\bibfield  {journal} {\bibinfo  {journal} {Phys. Rev. A}\ }\textbf {\bibinfo
			{volume} {70}},\ \bibinfo {pages} {052328} (\bibinfo {year}
		{2004})}\BibitemShut {NoStop}%
	\bibitem [{\citenamefont {Bravyi}\ and\ \citenamefont
		{Kitaev}(2005)}]{bravyi2005universal}%
	\BibitemOpen
	\bibfield  {author} {\bibinfo {author} {\bibfnamefont {S.}~\bibnamefont
			{Bravyi}}\ and\ \bibinfo {author} {\bibfnamefont {A.}~\bibnamefont
			{Kitaev}},\ }\bibfield  {title} {\bibinfo {title} {Universal quantum
			computation with ideal {Clifford} gates and noisy ancillas},\ }\href
	{https://doi.org/10.1103/PhysRevA.71.022316} {\bibfield  {journal} {\bibinfo
			{journal} {Phys. Rev. A}\ }\textbf {\bibinfo {volume} {71}},\ \bibinfo
		{pages} {022316} (\bibinfo {year} {2005})}\BibitemShut {NoStop}%
	\bibitem [{\citenamefont {Veitch}\ \emph {et~al.}(2014)\citenamefont {Veitch},
		\citenamefont {Mousavian}, \citenamefont {Gottesman},\ and\ \citenamefont
		{Emerson}}]{veitch2014resource}%
	\BibitemOpen
	\bibfield  {author} {\bibinfo {author} {\bibfnamefont {V.}~\bibnamefont
			{Veitch}}, \bibinfo {author} {\bibfnamefont {S.~A.~H.}\ \bibnamefont
			{Mousavian}}, \bibinfo {author} {\bibfnamefont {D.}~\bibnamefont
			{Gottesman}},\ and\ \bibinfo {author} {\bibfnamefont {J.}~\bibnamefont
			{Emerson}},\ }\bibfield  {title} {\bibinfo {title} {The resource theory of
			stabilizer quantum computation},\ }\href
	{https://doi.org/10.1088/1367-2630/16/1/013009} {\bibfield  {journal}
		{\bibinfo  {journal} {New J. Phys.}\ }\textbf {\bibinfo {volume} {16}},\
		\bibinfo {pages} {013009} (\bibinfo {year} {2014})}\BibitemShut {NoStop}%
	\bibitem [{\citenamefont {Chitambar}\ and\ \citenamefont
		{Gour}(2019)}]{chitambar2019quantum}%
	\BibitemOpen
	\bibfield  {author} {\bibinfo {author} {\bibfnamefont {E.}~\bibnamefont
			{Chitambar}}\ and\ \bibinfo {author} {\bibfnamefont {G.}~\bibnamefont
			{Gour}},\ }\bibfield  {title} {\bibinfo {title} {Quantum resource theories},\
	}\href {https://doi.org/10.1103/RevModPhys.91.025001} {\bibfield  {journal}
		{\bibinfo  {journal} {Rev. Mod. Phys.}\ }\textbf {\bibinfo {volume} {91}},\
		\bibinfo {pages} {025001} (\bibinfo {year} {2019})}\BibitemShut {NoStop}%
	\bibitem [{\citenamefont {Bravyi}\ \emph {et~al.}(2016)\citenamefont {Bravyi},
		\citenamefont {Smith},\ and\ \citenamefont {Smolin}}]{bravyi2016trading}%
	\BibitemOpen
	\bibfield  {author} {\bibinfo {author} {\bibfnamefont {S.}~\bibnamefont
			{Bravyi}}, \bibinfo {author} {\bibfnamefont {G.}~\bibnamefont {Smith}},\ and\
		\bibinfo {author} {\bibfnamefont {J.~A.}\ \bibnamefont {Smolin}},\ }\bibfield
	{title} {\bibinfo {title} {Trading classical and quantum computational
			resources},\ }\href {https://doi.org/10.1103/PhysRevX.6.021043} {\bibfield
		{journal} {\bibinfo  {journal} {Phys. Rev. X}\ }\textbf {\bibinfo {volume}
			{6}},\ \bibinfo {pages} {021043} (\bibinfo {year} {2016})}\BibitemShut
	{NoStop}%
	\bibitem [{\citenamefont {Gross}\ \emph {et~al.}(2021)\citenamefont {Gross},
		\citenamefont {Nezami},\ and\ \citenamefont {Walter}}]{gross2021schur}%
	\BibitemOpen
	\bibfield  {author} {\bibinfo {author} {\bibfnamefont {D.}~\bibnamefont
			{Gross}}, \bibinfo {author} {\bibfnamefont {S.}~\bibnamefont {Nezami}},\ and\
		\bibinfo {author} {\bibfnamefont {M.}~\bibnamefont {Walter}},\ }\bibfield
	{title} {\bibinfo {title} {{Schur--Weyl} duality for the {Clifford} group
			with applications: Property testing, a robust {Hudson} theorem, and {de
				Finetti} representations},\ }\href
	{https://doi.org/10.1007/s00220-021-04118-7} {\bibfield  {journal} {\bibinfo
			{journal} {Commun. Math. Phys.}\ }\textbf {\bibinfo {volume} {385}},\
		\bibinfo {pages} {1325} (\bibinfo {year} {2021})}\BibitemShut {NoStop}%
	\bibitem [{\citenamefont {Leone}\ \emph {et~al.}(2022)\citenamefont {Leone},
		\citenamefont {Oliviero},\ and\ \citenamefont
		{Hamma}}]{leone2022stabilizerrenyientropy}%
	\BibitemOpen
	\bibfield  {author} {\bibinfo {author} {\bibfnamefont {L.}~\bibnamefont
			{Leone}}, \bibinfo {author} {\bibfnamefont {S.~F.~E.}\ \bibnamefont
			{Oliviero}},\ and\ \bibinfo {author} {\bibfnamefont {A.}~\bibnamefont
			{Hamma}},\ }\bibfield  {title} {\bibinfo {title} {Stabilizer r\'enyi
			entropy},\ }\href {https://doi.org/10.1103/PhysRevLett.128.050402} {\bibfield
		{journal} {\bibinfo  {journal} {Phys. Rev. Lett.}\ }\textbf {\bibinfo
			{volume} {128}},\ \bibinfo {pages} {050402} (\bibinfo {year}
		{2022})}\BibitemShut {NoStop}%
	\bibitem [{\citenamefont {Haug}\ and\ \citenamefont
		{Kim}(2023)}]{haug2023scalable}%
	\BibitemOpen
	\bibfield  {author} {\bibinfo {author} {\bibfnamefont {T.}~\bibnamefont
			{Haug}}\ and\ \bibinfo {author} {\bibfnamefont {M.}~\bibnamefont {Kim}},\
	}\bibfield  {title} {\bibinfo {title} {Scalable measures of magic resource
			for quantum computers},\ }\href {https://doi.org/10.1103/PRXQuantum.4.010301}
	{\bibfield  {journal} {\bibinfo  {journal} {PRX Quantum}\ }\textbf {\bibinfo
			{volume} {4}},\ \bibinfo {pages} {010301} (\bibinfo {year}
		{2023})}\BibitemShut {NoStop}%
	\bibitem [{\citenamefont {Haug}\ and\ \citenamefont
		{Piroli}(2023{\natexlab{a}})}]{haug2023quantifying}%
	\BibitemOpen
	\bibfield  {author} {\bibinfo {author} {\bibfnamefont {T.}~\bibnamefont
			{Haug}}\ and\ \bibinfo {author} {\bibfnamefont {L.}~\bibnamefont {Piroli}},\
	}\bibfield  {title} {\bibinfo {title} {Quantifying nonstabilizerness of
			matrix product states},\ }\href {https://doi.org/10.1103/PhysRevB.107.035148}
	{\bibfield  {journal} {\bibinfo  {journal} {Phys. Rev. B}\ }\textbf {\bibinfo
			{volume} {107}},\ \bibinfo {pages} {035148} (\bibinfo {year}
		{2023}{\natexlab{a}})}\BibitemShut {NoStop}%
	\bibitem [{\citenamefont {Haug}\ and\ \citenamefont
		{Piroli}(2023{\natexlab{b}})}]{haug2023stabilizer}%
	\BibitemOpen
	\bibfield  {author} {\bibinfo {author} {\bibfnamefont {T.}~\bibnamefont
			{Haug}}\ and\ \bibinfo {author} {\bibfnamefont {L.}~\bibnamefont {Piroli}},\
	}\bibfield  {title} {\bibinfo {title} {Stabilizer entropies and
			nonstabilizerness monotones},\ }\href
	{https://doi.org/10.22331/q-2023-08-28-1092} {\bibfield  {journal} {\bibinfo
			{journal} {{Quantum}}\ }\textbf {\bibinfo {volume} {7}},\ \bibinfo {pages}
		{1092} (\bibinfo {year} {2023}{\natexlab{b}})}\BibitemShut {NoStop}%
	\bibitem [{\citenamefont {Leone}\ and\ \citenamefont
		{Bittel}(2024)}]{leone2024stabilizer}%
	\BibitemOpen
	\bibfield  {author} {\bibinfo {author} {\bibfnamefont {L.}~\bibnamefont
			{Leone}}\ and\ \bibinfo {author} {\bibfnamefont {L.}~\bibnamefont {Bittel}},\
	}\bibfield  {title} {\bibinfo {title} {Stabilizer entropies are monotones for
			magic-state resource theory},\ }\href
	{https://doi.org/10.1103/PhysRevA.110.L040403} {\bibfield  {journal}
		{\bibinfo  {journal} {Phys. Rev. A}\ }\textbf {\bibinfo {volume} {110}},\
		\bibinfo {pages} {L040403} (\bibinfo {year} {2024})}\BibitemShut {NoStop}%
	\bibitem [{\citenamefont {Turkeshi}\ \emph {et~al.}(2025)\citenamefont
		{Turkeshi}, \citenamefont {Tirrito},\ and\ \citenamefont
		{Sierant}}]{turkeshi2024magic}%
	\BibitemOpen
	\bibfield  {author} {\bibinfo {author} {\bibfnamefont {X.}~\bibnamefont
			{Turkeshi}}, \bibinfo {author} {\bibfnamefont {E.}~\bibnamefont {Tirrito}},\
		and\ \bibinfo {author} {\bibfnamefont {P.}~\bibnamefont {Sierant}},\
	}\bibfield  {title} {\bibinfo {title} {Magic spreading in random quantum
			circuits},\ }\href {https://doi.org/10.1038/s41467-025-57704-x} {\bibfield
		{journal} {\bibinfo  {journal} {Nat Commun}\ }\textbf {\bibinfo {volume}
			{16}},\ \bibinfo {pages} {2575} (\bibinfo {year} {2025})}\BibitemShut
	{NoStop}%
	\bibitem [{\citenamefont {Yoganathan}\ \emph {et~al.}(2019)\citenamefont
		{Yoganathan}, \citenamefont {Jozsa},\ and\ \citenamefont
		{Strelchuk}}]{yoganathan2019quantum}%
	\BibitemOpen
	\bibfield  {author} {\bibinfo {author} {\bibfnamefont {M.}~\bibnamefont
			{Yoganathan}}, \bibinfo {author} {\bibfnamefont {R.}~\bibnamefont {Jozsa}},\
		and\ \bibinfo {author} {\bibfnamefont {S.}~\bibnamefont {Strelchuk}},\
	}\bibfield  {title} {\bibinfo {title} {Quantum advantage of unitary
			{Clifford} circuits with magic state inputs},\ }\href
	{https://doi.org/10.1098/rspa.2018.0427} {\bibfield  {journal} {\bibinfo
			{journal} {Proc. Math. Phys. Eng. Sci.}\ }\textbf {\bibinfo {volume} {475}},\
		\bibinfo {pages} {20180427} (\bibinfo {year} {2019})}\BibitemShut {NoStop}%
	\bibitem [{\citenamefont {White}\ \emph {et~al.}(2021)\citenamefont {White},
		\citenamefont {Cao},\ and\ \citenamefont {Swingle}}]{white2021conformal}%
	\BibitemOpen
	\bibfield  {author} {\bibinfo {author} {\bibfnamefont {C.~D.}\ \bibnamefont
			{White}}, \bibinfo {author} {\bibfnamefont {C.}~\bibnamefont {Cao}},\ and\
		\bibinfo {author} {\bibfnamefont {B.}~\bibnamefont {Swingle}},\ }\bibfield
	{title} {\bibinfo {title} {Conformal field theories are magical},\ }\href
	{https://doi.org/10.1103/PhysRevB.103.075145} {\bibfield  {journal} {\bibinfo
			{journal} {Phys. Rev. B}\ }\textbf {\bibinfo {volume} {103}},\ \bibinfo
		{pages} {075145} (\bibinfo {year} {2021})}\BibitemShut {NoStop}%
	\bibitem [{\citenamefont {Lami}\ and\ \citenamefont
		{Collura}(2023)}]{lami2023nonstabilizerness}%
	\BibitemOpen
	\bibfield  {author} {\bibinfo {author} {\bibfnamefont {G.}~\bibnamefont
			{Lami}}\ and\ \bibinfo {author} {\bibfnamefont {M.}~\bibnamefont {Collura}},\
	}\bibfield  {title} {\bibinfo {title} {Nonstabilizerness via perfect pauli
			sampling of matrix product states},\ }\href
	{https://doi.org/10.1103/PhysRevLett.131.180401} {\bibfield  {journal}
		{\bibinfo  {journal} {Phys. Rev. Lett.}\ }\textbf {\bibinfo {volume} {131}},\
		\bibinfo {pages} {180401} (\bibinfo {year} {2023})}\BibitemShut {NoStop}%
	\bibitem [{\citenamefont {Chen}\ \emph {et~al.}(2024)\citenamefont {Chen},
		\citenamefont {Garcia}, \citenamefont {Bu},\ and\ \citenamefont
		{Jaffe}}]{chen2024magic}%
	\BibitemOpen
	\bibfield  {author} {\bibinfo {author} {\bibfnamefont {L.}~\bibnamefont
			{Chen}}, \bibinfo {author} {\bibfnamefont {R.~J.}\ \bibnamefont {Garcia}},
		\bibinfo {author} {\bibfnamefont {K.}~\bibnamefont {Bu}},\ and\ \bibinfo
		{author} {\bibfnamefont {A.}~\bibnamefont {Jaffe}},\ }\bibfield  {title}
	{\bibinfo {title} {Magic of random matrix product states},\ }\href
	{https://doi.org/10.1103/PhysRevB.109.174207} {\bibfield  {journal} {\bibinfo
			{journal} {Phys. Rev. B}\ }\textbf {\bibinfo {volume} {109}},\ \bibinfo
		{pages} {174207} (\bibinfo {year} {2024})}\BibitemShut {NoStop}%
	\bibitem [{\citenamefont {Tarabunga}\ \emph {et~al.}(2023)\citenamefont
		{Tarabunga}, \citenamefont {Tirrito}, \citenamefont {Chanda},\ and\
		\citenamefont {Dalmonte}}]{tarabunga2023manybodymagic}%
	\BibitemOpen
	\bibfield  {author} {\bibinfo {author} {\bibfnamefont {P.~S.}\ \bibnamefont
			{Tarabunga}}, \bibinfo {author} {\bibfnamefont {E.}~\bibnamefont {Tirrito}},
		\bibinfo {author} {\bibfnamefont {T.}~\bibnamefont {Chanda}},\ and\ \bibinfo
		{author} {\bibfnamefont {M.}~\bibnamefont {Dalmonte}},\ }\bibfield  {title}
	{\bibinfo {title} {Many-body magic via {Pauli-Markov} chains---from
			criticality to gauge theories},\ }\href
	{https://doi.org/10.1103/PRXQuantum.4.040317} {\bibfield  {journal} {\bibinfo
			{journal} {PRX Quantum}\ }\textbf {\bibinfo {volume} {4}},\ \bibinfo {pages}
		{040317} (\bibinfo {year} {2023})}\BibitemShut {NoStop}%
	\bibitem [{\citenamefont {Tarabunga}\ \emph {et~al.}(2024)\citenamefont
		{Tarabunga}, \citenamefont {Tirrito}, \citenamefont {Ba\~nuls},\ and\
		\citenamefont {Dalmonte}}]{tarabunga2024nonstabilizerness}%
	\BibitemOpen
	\bibfield  {author} {\bibinfo {author} {\bibfnamefont {P.~S.}\ \bibnamefont
			{Tarabunga}}, \bibinfo {author} {\bibfnamefont {E.}~\bibnamefont {Tirrito}},
		\bibinfo {author} {\bibfnamefont {M.~C.}\ \bibnamefont {Ba\~nuls}},\ and\
		\bibinfo {author} {\bibfnamefont {M.}~\bibnamefont {Dalmonte}},\ }\bibfield
	{title} {\bibinfo {title} {Nonstabilizerness via matrix product states in the
			pauli basis},\ }\href {https://doi.org/10.1103/PhysRevLett.133.010601}
	{\bibfield  {journal} {\bibinfo  {journal} {Phys. Rev. Lett.}\ }\textbf
		{\bibinfo {volume} {133}},\ \bibinfo {pages} {010601} (\bibinfo {year}
		{2024})}\BibitemShut {NoStop}%
	\bibitem [{\citenamefont {Tarabunga}\ and\ \citenamefont
		{Castelnovo}(2024)}]{tarabunga2024magic}%
	\BibitemOpen
	\bibfield  {author} {\bibinfo {author} {\bibfnamefont {P.~S.}\ \bibnamefont
			{Tarabunga}}\ and\ \bibinfo {author} {\bibfnamefont {C.}~\bibnamefont
			{Castelnovo}},\ }\bibfield  {title} {\bibinfo {title} {Magic in generalized
			{Rokhsar-Kivelson} wavefunctions},\ }\href
	{https://doi.org/10.22331/q-2024-05-14-1347} {\bibfield  {journal} {\bibinfo
			{journal} {Quantum}\ }\textbf {\bibinfo {volume} {8}},\ \bibinfo {pages}
		{1347} (\bibinfo {year} {2024})}\BibitemShut {NoStop}%
	\bibitem [{\citenamefont {Tarabunga}(2024)}]{tarabunga2024critical}%
	\BibitemOpen
	\bibfield  {author} {\bibinfo {author} {\bibfnamefont {P.~S.}\ \bibnamefont
			{Tarabunga}},\ }\bibfield  {title} {\bibinfo {title} {Critical behaviors of
			non-stabilizerness in quantum spin chains},\ }\href
	{https://doi.org/10.22331/q-2024-07-17-1413} {\bibfield  {journal} {\bibinfo
			{journal} {Quantum}\ }\textbf {\bibinfo {volume} {8}},\ \bibinfo {pages}
		{1413} (\bibinfo {year} {2024})}\BibitemShut {NoStop}%
	\bibitem [{\citenamefont {Bejan}\ \emph {et~al.}(2024)\citenamefont {Bejan},
		\citenamefont {McLauchlan},\ and\ \citenamefont
		{B\'eri}}]{bejan2024dynamical}%
	\BibitemOpen
	\bibfield  {author} {\bibinfo {author} {\bibfnamefont {M.}~\bibnamefont
			{Bejan}}, \bibinfo {author} {\bibfnamefont {C.}~\bibnamefont {McLauchlan}},\
		and\ \bibinfo {author} {\bibfnamefont {B.}~\bibnamefont {B\'eri}},\
	}\bibfield  {title} {\bibinfo {title} {Dynamical magic transitions in
			monitored {Clifford+$T$} circuits},\ }\href
	{https://doi.org/10.1103/PRXQuantum.5.030332} {\bibfield  {journal} {\bibinfo
			{journal} {PRX Quantum}\ }\textbf {\bibinfo {volume} {5}},\ \bibinfo {pages}
		{030332} (\bibinfo {year} {2024})}\BibitemShut {NoStop}%
	\bibitem [{\citenamefont {Bluvstein}\ \emph {et~al.}(2024)\citenamefont
		{Bluvstein}, \citenamefont {Evered}, \citenamefont {Geim}, \citenamefont
		{Li}, \citenamefont {Zhou}, \citenamefont {Manovitz}, \citenamefont {Ebadi},
		\citenamefont {Cain}, \citenamefont {Kalinowski}, \citenamefont {Hangleiter}
		\emph {et~al.}}]{bluvstein2024logical}%
	\BibitemOpen
	\bibfield  {author} {\bibinfo {author} {\bibfnamefont {D.}~\bibnamefont
			{Bluvstein}}, \bibinfo {author} {\bibfnamefont {S.~J.}\ \bibnamefont
			{Evered}}, \bibinfo {author} {\bibfnamefont {A.~A.}\ \bibnamefont {Geim}},
		\bibinfo {author} {\bibfnamefont {S.~H.}\ \bibnamefont {Li}}, \bibinfo
		{author} {\bibfnamefont {H.}~\bibnamefont {Zhou}}, \bibinfo {author}
		{\bibfnamefont {T.}~\bibnamefont {Manovitz}}, \bibinfo {author}
		{\bibfnamefont {S.}~\bibnamefont {Ebadi}}, \bibinfo {author} {\bibfnamefont
			{M.}~\bibnamefont {Cain}}, \bibinfo {author} {\bibfnamefont {M.}~\bibnamefont
			{Kalinowski}}, \bibinfo {author} {\bibfnamefont {D.}~\bibnamefont
			{Hangleiter}}, \emph {et~al.},\ }\bibfield  {title} {\bibinfo {title}
		{Logical quantum processor based on reconfigurable atom arrays},\ }\href
	{https://doi.org/10.1038/s41586-023-06927-3} {\bibfield  {journal} {\bibinfo
			{journal} {Nature}\ }\textbf {\bibinfo {volume} {626}},\ \bibinfo {pages}
		{58} (\bibinfo {year} {2024})}\BibitemShut {NoStop}%
	\bibitem [{\citenamefont {Oliviero}\ \emph {et~al.}(2022)\citenamefont
		{Oliviero}, \citenamefont {Leone}, \citenamefont {Hamma},\ and\ \citenamefont
		{Lloyd}}]{oliviero2022measuring}%
	\BibitemOpen
	\bibfield  {author} {\bibinfo {author} {\bibfnamefont {S.~F.}\ \bibnamefont
			{Oliviero}}, \bibinfo {author} {\bibfnamefont {L.}~\bibnamefont {Leone}},
		\bibinfo {author} {\bibfnamefont {A.}~\bibnamefont {Hamma}},\ and\ \bibinfo
		{author} {\bibfnamefont {S.}~\bibnamefont {Lloyd}},\ }\bibfield  {title}
	{\bibinfo {title} {Measuring magic on a quantum processor},\ }\href
	{https://doi.org/10.1038/s41534-022-00666-5} {\bibfield  {journal} {\bibinfo
			{journal} {npj Quantum Inf.}\ }\textbf {\bibinfo {volume} {8}},\ \bibinfo
		{pages} {148} (\bibinfo {year} {2022})}\BibitemShut {NoStop}%
	\bibitem [{\citenamefont {Haug}\ \emph {et~al.}(2024)\citenamefont {Haug},
		\citenamefont {Lee},\ and\ \citenamefont {Kim}}]{haug2024efficient}%
	\BibitemOpen
	\bibfield  {author} {\bibinfo {author} {\bibfnamefont {T.}~\bibnamefont
			{Haug}}, \bibinfo {author} {\bibfnamefont {S.}~\bibnamefont {Lee}},\ and\
		\bibinfo {author} {\bibfnamefont {M.~S.}\ \bibnamefont {Kim}},\ }\bibfield
	{title} {\bibinfo {title} {Efficient quantum algorithms for stabilizer
			entropies},\ }\href {https://doi.org/10.1103/PhysRevLett.132.240602}
	{\bibfield  {journal} {\bibinfo  {journal} {Phys. Rev. Lett.}\ }\textbf
		{\bibinfo {volume} {132}},\ \bibinfo {pages} {240602} (\bibinfo {year}
		{2024})}\BibitemShut {NoStop}%
	\bibitem [{\citenamefont {Niroula}\ \emph {et~al.}(2024)\citenamefont
		{Niroula}, \citenamefont {White}, \citenamefont {Wang}, \citenamefont
		{Johri}, \citenamefont {Zhu}, \citenamefont {Monroe}, \citenamefont {Noel},\
		and\ \citenamefont {Gullans}}]{niroula2024phase}%
	\BibitemOpen
	\bibfield  {author} {\bibinfo {author} {\bibfnamefont {P.}~\bibnamefont
			{Niroula}}, \bibinfo {author} {\bibfnamefont {C.~D.}\ \bibnamefont {White}},
		\bibinfo {author} {\bibfnamefont {Q.}~\bibnamefont {Wang}}, \bibinfo {author}
		{\bibfnamefont {S.}~\bibnamefont {Johri}}, \bibinfo {author} {\bibfnamefont
			{D.}~\bibnamefont {Zhu}}, \bibinfo {author} {\bibfnamefont {C.}~\bibnamefont
			{Monroe}}, \bibinfo {author} {\bibfnamefont {C.}~\bibnamefont {Noel}},\ and\
		\bibinfo {author} {\bibfnamefont {M.~J.}\ \bibnamefont {Gullans}},\
	}\bibfield  {title} {\bibinfo {title} {Phase transition in magic with random
			quantum circuits},\ }\bibfield  {journal} {\bibinfo  {journal} {Nat. Phys.}\
	}\href {https://doi.org/10.1038/s41567-024-02637-3}
	{10.1038/s41567-024-02637-3} (\bibinfo {year} {2024})\BibitemShut {NoStop}%
	\bibitem [{\citenamefont {Haferkamp}\ \emph {et~al.}(2023)\citenamefont
		{Haferkamp}, \citenamefont {Montealegre-Mora}, \citenamefont {Heinrich},
		\citenamefont {Eisert}, \citenamefont {Gross},\ and\ \citenamefont
		{Roth}}]{haferkamp2023efficient}%
	\BibitemOpen
	\bibfield  {author} {\bibinfo {author} {\bibfnamefont {J.}~\bibnamefont
			{Haferkamp}}, \bibinfo {author} {\bibfnamefont {F.}~\bibnamefont
			{Montealegre-Mora}}, \bibinfo {author} {\bibfnamefont {M.}~\bibnamefont
			{Heinrich}}, \bibinfo {author} {\bibfnamefont {J.}~\bibnamefont {Eisert}},
		\bibinfo {author} {\bibfnamefont {D.}~\bibnamefont {Gross}},\ and\ \bibinfo
		{author} {\bibfnamefont {I.}~\bibnamefont {Roth}},\ }\bibfield  {title}
	{\bibinfo {title} {Efficient unitary designs with a system-size independent
			number of non-{Clifford} gates},\ }\href
	{https://doi.org/10.1007/s00220-022-04507-6} {\bibfield  {journal} {\bibinfo
			{journal} {Commun. Math. Phys.}\ }\textbf {\bibinfo {volume} {397}},\
		\bibinfo {pages} {995} (\bibinfo {year} {2023})}\BibitemShut {NoStop}%
	\bibitem [{Note1()}]{Note1}%
	\BibitemOpen
	\bibinfo {note} {The choice of T-gates on the first qubit is equivalent with
		any other choice $G_i = L_i T R_i$ for any Cliffords $L_i, R_i \in \protect
		\mathcal {C}_N$ (where $\protect \mathcal {C}_N$ denotes the set of Clifford
		unitarys on $N$ qubits) because they could be absorbed into the adjacent
		random Clifford circuits $C_i$ and $C_{i+1}$ respectively. In particular
		(because $\protect \mathtt {SWAP}\in \protect \mathcal {C}_2$) these circuits
		are equivalent to circuits with T-gates applied to any sequence of
		qubits.}\BibitemShut {Stop}%
	\bibitem [{\citenamefont {Beverland}\ \emph {et~al.}(2020)\citenamefont
		{Beverland}, \citenamefont {Campbell}, \citenamefont {Howard},\ and\
		\citenamefont {Kliuchnikov}}]{beverland2020lower}%
	\BibitemOpen
	\bibfield  {author} {\bibinfo {author} {\bibfnamefont {M.}~\bibnamefont
			{Beverland}}, \bibinfo {author} {\bibfnamefont {E.}~\bibnamefont {Campbell}},
		\bibinfo {author} {\bibfnamefont {M.}~\bibnamefont {Howard}},\ and\ \bibinfo
		{author} {\bibfnamefont {V.}~\bibnamefont {Kliuchnikov}},\ }\bibfield
	{title} {\bibinfo {title} {Lower bounds on the non-{Clifford} resources for
			quantum computations},\ }\href {https://doi.org/10.1088/2058-9565/ab8963}
	{\bibfield  {journal} {\bibinfo  {journal} {Quantum Sci. Technol.}\ }\textbf
		{\bibinfo {volume} {5}},\ \bibinfo {pages} {035009} (\bibinfo {year}
		{2020})}\BibitemShut {NoStop}%
	\bibitem [{\citenamefont {Jiang}\ and\ \citenamefont
		{Wang}(2023)}]{jiang2023lower}%
	\BibitemOpen
	\bibfield  {author} {\bibinfo {author} {\bibfnamefont {J.}~\bibnamefont
			{Jiang}}\ and\ \bibinfo {author} {\bibfnamefont {X.}~\bibnamefont {Wang}},\
	}\bibfield  {title} {\bibinfo {title} {Lower bound for the $t$ count via
			unitary stabilizer nullity},\ }\href
	{https://doi.org/10.1103/PhysRevApplied.19.034052} {\bibfield  {journal}
		{\bibinfo  {journal} {Phys. Rev. Appl.}\ }\textbf {\bibinfo {volume} {19}},\
		\bibinfo {pages} {034052} (\bibinfo {year} {2023})}\BibitemShut {NoStop}%
	\bibitem [{\citenamefont {Amy}\ and\ \citenamefont
		{Mosca}(2019)}]{amy2019tcount}%
	\BibitemOpen
	\bibfield  {author} {\bibinfo {author} {\bibfnamefont {M.}~\bibnamefont
			{Amy}}\ and\ \bibinfo {author} {\bibfnamefont {M.}~\bibnamefont {Mosca}},\
	}\bibfield  {title} {\bibinfo {title} {T-count optimization and {Reed-Muller}
			codes},\ }\href {https://doi.org/10.1109/TIT.2019.2906374} {\bibfield
		{journal} {\bibinfo  {journal} {IEEE Trans. Inform. Theory}\ }\textbf
		{\bibinfo {volume} {65}},\ \bibinfo {pages} {4771} (\bibinfo {year}
		{2019})}\BibitemShut {NoStop}%
	\bibitem [{Note2()}]{Note2}%
	\BibitemOpen
	\bibinfo {note} {The error is exponentially small in $N$: by our
		probabilistic argument below, $\protect \text {Pr}(Z_1\in \protect \mathcal
		{S})\leq 2^{N}/(4^N-1)$.}\BibitemShut {Stop}%
	\bibitem [{\citenamefont {Masot-Llima}\ and\ \citenamefont
		{Garcia-Saez}(2024)}]{masotllima2024stabilizer}%
	\BibitemOpen
	\bibfield  {author} {\bibinfo {author} {\bibfnamefont {S.}~\bibnamefont
			{Masot-Llima}}\ and\ \bibinfo {author} {\bibfnamefont {A.}~\bibnamefont
			{Garcia-Saez}},\ }\bibfield  {title} {\bibinfo {title} {Stabilizer tensor
			networks: Universal quantum simulator on a basis of stabilizer states},\
	}\href {https://doi.org/10.1103/PhysRevLett.133.230601} {\bibfield  {journal}
		{\bibinfo  {journal} {Phys. Rev. Lett.}\ }\textbf {\bibinfo {volume} {133}},\
		\bibinfo {pages} {230601} (\bibinfo {year} {2024})}\BibitemShut {NoStop}%
	\bibitem [{\citenamefont {Qian}\ \emph {et~al.}(2024)\citenamefont {Qian},
		\citenamefont {Huang},\ and\ \citenamefont {Qin}}]{qian2024augmenting}%
	\BibitemOpen
	\bibfield  {author} {\bibinfo {author} {\bibfnamefont {X.}~\bibnamefont
			{Qian}}, \bibinfo {author} {\bibfnamefont {J.}~\bibnamefont {Huang}},\ and\
		\bibinfo {author} {\bibfnamefont {M.}~\bibnamefont {Qin}},\ }\bibfield
	{title} {\bibinfo {title} {Augmenting density matrix renormalization group
			with {Clifford} circuits},\ }\href
	{https://doi.org/10.1103/PhysRevLett.133.190402} {\bibfield  {journal}
		{\bibinfo  {journal} {Phys. Rev. Lett.}\ }\textbf {\bibinfo {volume} {133}},\
		\bibinfo {pages} {190402} (\bibinfo {year} {2024})}\BibitemShut {NoStop}%
	\bibitem [{\citenamefont {Huang}\ \emph {et~al.}(2025)\citenamefont {Huang},
		\citenamefont {Qian},\ and\ \citenamefont
		{Qin}}]{huang2024nonstabilizerness}%
	\BibitemOpen
	\bibfield  {author} {\bibinfo {author} {\bibfnamefont {J.}~\bibnamefont
			{Huang}}, \bibinfo {author} {\bibfnamefont {X.}~\bibnamefont {Qian}},\ and\
		\bibinfo {author} {\bibfnamefont {M.}~\bibnamefont {Qin}},\ }\bibfield
	{title} {\bibinfo {title} {Nonstabilizerness entanglement entropy: A measure
			of hardness in the classical simulation of quantum many-body systems with
			tensor network states},\ }\href {https://doi.org/10.1103/gxdn-zwrw}
	{\bibfield  {journal} {\bibinfo  {journal} {Phys. Rev. A}\ }\textbf {\bibinfo
			{volume} {112}},\ \bibinfo {pages} {012425} (\bibinfo {year}
		{2025})}\BibitemShut {NoStop}%
	\bibitem [{\citenamefont {Chamon}\ \emph {et~al.}(2014)\citenamefont {Chamon},
		\citenamefont {Hamma},\ and\ \citenamefont {Mucciolo}}]{chamon2014emergent}%
	\BibitemOpen
	\bibfield  {author} {\bibinfo {author} {\bibfnamefont {C.}~\bibnamefont
			{Chamon}}, \bibinfo {author} {\bibfnamefont {A.}~\bibnamefont {Hamma}},\ and\
		\bibinfo {author} {\bibfnamefont {E.~R.}\ \bibnamefont {Mucciolo}},\
	}\bibfield  {title} {\bibinfo {title} {Emergent irreversibility and
			entanglement spectrum statistics},\ }\href
	{https://doi.org/10.1103/PhysRevLett.112.240501} {\bibfield  {journal}
		{\bibinfo  {journal} {Phys. Rev. Lett.}\ }\textbf {\bibinfo {volume} {112}},\
		\bibinfo {pages} {240501} (\bibinfo {year} {2014})}\BibitemShut {NoStop}%
	\bibitem [{\citenamefont {Shaffer}\ \emph {et~al.}(2014)\citenamefont
		{Shaffer}, \citenamefont {Chamon}, \citenamefont {Hamma},\ and\ \citenamefont
		{Mucciolo}}]{shaffer2014irreversibility}%
	\BibitemOpen
	\bibfield  {author} {\bibinfo {author} {\bibfnamefont {D.}~\bibnamefont
			{Shaffer}}, \bibinfo {author} {\bibfnamefont {C.}~\bibnamefont {Chamon}},
		\bibinfo {author} {\bibfnamefont {A.}~\bibnamefont {Hamma}},\ and\ \bibinfo
		{author} {\bibfnamefont {E.~R.}\ \bibnamefont {Mucciolo}},\ }\bibfield
	{title} {\bibinfo {title} {Irreversibility and entanglement spectrum
			statistics in quantum circuits},\ }\href
	{https://doi.org/10.1088/1742-5468/2014/12/P12007} {\bibfield  {journal}
		{\bibinfo  {journal} {J. Stat. Mech. Theory Exp.}\ }\textbf {\bibinfo
			{volume} {2014}},\ \bibinfo {pages} {P12007} (\bibinfo {year}
		{2014})}\BibitemShut {NoStop}%
	\bibitem [{\citenamefont {True}\ and\ \citenamefont
		{Hamma}(2022)}]{true2022transitions}%
	\BibitemOpen
	\bibfield  {author} {\bibinfo {author} {\bibfnamefont {S.}~\bibnamefont
			{True}}\ and\ \bibinfo {author} {\bibfnamefont {A.}~\bibnamefont {Hamma}},\
	}\bibfield  {title} {\bibinfo {title} {Transitions in entanglement complexity
			in random circuits},\ }\href {https://doi.org/10.22331/q-2022-09-22-818}
	{\bibfield  {journal} {\bibinfo  {journal} {{Quantum}}\ }\textbf {\bibinfo
			{volume} {6}},\ \bibinfo {pages} {818} (\bibinfo {year} {2022})}\BibitemShut
	{NoStop}%
	\bibitem [{\citenamefont {Mello}\ \emph {et~al.}(2024)\citenamefont {Mello},
		\citenamefont {Santini},\ and\ \citenamefont {Collura}}]{mello2024hybrid}%
	\BibitemOpen
	\bibfield  {author} {\bibinfo {author} {\bibfnamefont {A.~F.}\ \bibnamefont
			{Mello}}, \bibinfo {author} {\bibfnamefont {A.}~\bibnamefont {Santini}},\
		and\ \bibinfo {author} {\bibfnamefont {M.}~\bibnamefont {Collura}},\
	}\bibfield  {title} {\bibinfo {title} {Hybrid stabilizer matrix product
			operator},\ }\href {https://doi.org/10.1103/PhysRevLett.133.150604}
	{\bibfield  {journal} {\bibinfo  {journal} {Phys. Rev. Lett.}\ }\textbf
		{\bibinfo {volume} {133}},\ \bibinfo {pages} {150604} (\bibinfo {year}
		{2024})}\BibitemShut {NoStop}%
	\bibitem [{\citenamefont {Mello}\ \emph {et~al.}(2025)\citenamefont {Mello},
		\citenamefont {Santini}, \citenamefont {Lami}, \citenamefont {De~Nardis},\
		and\ \citenamefont {Collura}}]{mello2024clifford}%
	\BibitemOpen
	\bibfield  {author} {\bibinfo {author} {\bibfnamefont {A.~F.}\ \bibnamefont
			{Mello}}, \bibinfo {author} {\bibfnamefont {A.}~\bibnamefont {Santini}},
		\bibinfo {author} {\bibfnamefont {G.}~\bibnamefont {Lami}}, \bibinfo {author}
		{\bibfnamefont {J.}~\bibnamefont {De~Nardis}},\ and\ \bibinfo {author}
		{\bibfnamefont {M.}~\bibnamefont {Collura}},\ }\bibfield  {title} {\bibinfo
		{title} {Clifford dressed time-dependent variational principle},\ }\href
	{https://doi.org/10.1103/PhysRevLett.134.150403} {\bibfield  {journal}
		{\bibinfo  {journal} {Phys. Rev. Lett.}\ }\textbf {\bibinfo {volume} {134}},\
		\bibinfo {pages} {150403} (\bibinfo {year} {2025})}\BibitemShut {NoStop}%
	\bibitem [{\citenamefont {Qian}\ \emph {et~al.}(2025)\citenamefont {Qian},
		\citenamefont {Huang},\ and\ \citenamefont {Qin}}]{qian2024clifford}%
	\BibitemOpen
	\bibfield  {author} {\bibinfo {author} {\bibfnamefont {X.}~\bibnamefont
			{Qian}}, \bibinfo {author} {\bibfnamefont {J.}~\bibnamefont {Huang}},\ and\
		\bibinfo {author} {\bibfnamefont {M.}~\bibnamefont {Qin}},\ }\bibfield
	{title} {\bibinfo {title} {Clifford circuits augmented time-dependent
			variational principle},\ }\href
	{https://doi.org/10.1103/PhysRevLett.134.150404} {\bibfield  {journal}
		{\bibinfo  {journal} {Phys. Rev. Lett.}\ }\textbf {\bibinfo {volume} {134}},\
		\bibinfo {pages} {150404} (\bibinfo {year} {2025})}\BibitemShut {NoStop}%
	\bibitem [{\citenamefont {Nakhl}\ \emph {et~al.}(2025)\citenamefont {Nakhl},
		\citenamefont {Harper}, \citenamefont {West}, \citenamefont {Dowling},
		\citenamefont {Sevior}, \citenamefont {Quella},\ and\ \citenamefont
		{Usman}}]{nakhl2024stabilizer}%
	\BibitemOpen
	\bibfield  {author} {\bibinfo {author} {\bibfnamefont {A.~C.}\ \bibnamefont
			{Nakhl}}, \bibinfo {author} {\bibfnamefont {B.}~\bibnamefont {Harper}},
		\bibinfo {author} {\bibfnamefont {M.}~\bibnamefont {West}}, \bibinfo {author}
		{\bibfnamefont {N.}~\bibnamefont {Dowling}}, \bibinfo {author} {\bibfnamefont
			{M.}~\bibnamefont {Sevior}}, \bibinfo {author} {\bibfnamefont
			{T.}~\bibnamefont {Quella}},\ and\ \bibinfo {author} {\bibfnamefont
			{M.}~\bibnamefont {Usman}},\ }\bibfield  {title} {\bibinfo {title}
		{Stabilizer tensor networks with magic state injection},\ }\href
	{https://doi.org/10.1103/PhysRevLett.134.190602} {\bibfield  {journal}
		{\bibinfo  {journal} {Phys. Rev. Lett.}\ }\textbf {\bibinfo {volume} {134}},\
		\bibinfo {pages} {190602} (\bibinfo {year} {2025})}\BibitemShut {NoStop}%
	\bibitem [{\citenamefont {Liu}\ and\ \citenamefont
		{Clark}(2024)}]{liu2024classical}%
	\BibitemOpen
	\bibfield  {author} {\bibinfo {author} {\bibfnamefont {Z.}~\bibnamefont
			{Liu}}\ and\ \bibinfo {author} {\bibfnamefont {B.~K.}\ \bibnamefont
			{Clark}},\ }\bibfield  {title} {\bibinfo {title} {Classical simulability of
			{Clifford+T} circuits with {Clifford}-augmented matrix product states},\
	}\href@noop {} {\  (\bibinfo {year} {2024})},\ \Eprint
	{https://arxiv.org/abs/2412.17209} {arXiv:2412.17209} \BibitemShut {NoStop}%
	\bibitem [{Note3()}]{Note3}%
	\BibitemOpen
	\bibinfo {note} {Reference~\cite {lami2024quantum} mentions a similar
		numerical observation in its appendix.}\BibitemShut {Stop}%
	\bibitem [{\citenamefont {Suzuki}(1992)}]{suzuki1992general}%
	\BibitemOpen
	\bibfield  {author} {\bibinfo {author} {\bibfnamefont {M.}~\bibnamefont
			{Suzuki}},\ }\bibfield  {title} {\bibinfo {title} {General theory of
			higher-order decomposition of exponential operators and symplectic
			integrators},\ }\href {https://doi.org/10.1016/0375-9601(92)90335-J}
	{\bibfield  {journal} {\bibinfo  {journal} {Phys. Lett. A}\ }\textbf
		{\bibinfo {volume} {165}},\ \bibinfo {pages} {387} (\bibinfo {year}
		{1992})}\BibitemShut {NoStop}%
	\bibitem [{\citenamefont {Mele}(2024)}]{mele2024introduction}%
	\BibitemOpen
	\bibfield  {author} {\bibinfo {author} {\bibfnamefont {A.~A.}\ \bibnamefont
			{Mele}},\ }\bibfield  {title} {\bibinfo {title} {Introduction to haar measure
			tools in quantum information: A beginner's tutorial},\ }\href
	{https://doi.org/10.22331/q-2024-05-08-1340} {\bibfield  {journal} {\bibinfo
			{journal} {Quantum}\ }\textbf {\bibinfo {volume} {8}},\ \bibinfo {pages}
		{1340} (\bibinfo {year} {2024})}\BibitemShut {NoStop}%
	\bibitem [{Note4()}]{Note4}%
	\BibitemOpen
	\bibinfo {note} {Contributions to the state design that cannot be transformed
		into this form are suppressed exponentially in $N-t$.}\BibitemShut {Stop}%
	\bibitem [{Note5()}]{Note5}%
	\BibitemOpen
	\bibinfo {note} {By Remark 1 in Ref.~\cite {haferkamp2023efficient}, if $K^2$
		is non-Clifford, then $\phi _j$ may be sampled from just $\{\alpha , -\alpha
		\}$}\BibitemShut {NoStop}%
	\bibitem [{\citenamefont {Lami}\ \emph {et~al.}(2025)\citenamefont {Lami},
		\citenamefont {Haug},\ and\ \citenamefont {De~Nardis}}]{lami2024quantum}%
	\BibitemOpen
	\bibfield  {author} {\bibinfo {author} {\bibfnamefont {G.}~\bibnamefont
			{Lami}}, \bibinfo {author} {\bibfnamefont {T.}~\bibnamefont {Haug}},\ and\
		\bibinfo {author} {\bibfnamefont {J.}~\bibnamefont {De~Nardis}},\ }\bibfield
	{title} {\bibinfo {title} {Quantum state designs with {Clifford}-enhanced
			matrix product states},\ }\href {https://doi.org/10.1103/PRXQuantum.6.010345}
	{\bibfield  {journal} {\bibinfo  {journal} {PRX Quantum}\ }\textbf {\bibinfo
			{volume} {6}},\ \bibinfo {pages} {010345} (\bibinfo {year}
		{2025})}\BibitemShut {NoStop}%
	\bibitem [{\citenamefont {Shepherd}\ and\ \citenamefont
		{Bremner}(2009)}]{shepherd2009temporally}%
	\BibitemOpen
	\bibfield  {author} {\bibinfo {author} {\bibfnamefont {D.}~\bibnamefont
			{Shepherd}}\ and\ \bibinfo {author} {\bibfnamefont {M.~J.}\ \bibnamefont
			{Bremner}},\ }\bibfield  {title} {\bibinfo {title} {Temporally unstructured
			quantum computation},\ }\href {https://doi.org/10.1098/rspa.2008.0443}
	{\bibfield  {journal} {\bibinfo  {journal} {Proc. R. Soc. A.}\ }\textbf
		{\bibinfo {volume} {465}},\ \bibinfo {pages} {1413} (\bibinfo {year}
		{2009})}\BibitemShut {NoStop}%
	\bibitem [{\citenamefont {Bremner}\ \emph {et~al.}(2011)\citenamefont
		{Bremner}, \citenamefont {Jozsa},\ and\ \citenamefont
		{Shepherd}}]{bremner2011classical}%
	\BibitemOpen
	\bibfield  {author} {\bibinfo {author} {\bibfnamefont {M.~J.}\ \bibnamefont
			{Bremner}}, \bibinfo {author} {\bibfnamefont {R.}~\bibnamefont {Jozsa}},\
		and\ \bibinfo {author} {\bibfnamefont {D.~J.}\ \bibnamefont {Shepherd}},\
	}\bibfield  {title} {\bibinfo {title} {Classical simulation of commuting
			quantum computations implies collapse of the polynomial hierarchy},\ }\href
	{https://doi.org/10.1098/rspa.2010.0301} {\bibfield  {journal} {\bibinfo
			{journal} {Proc. R. Soc. A.}\ }\textbf {\bibinfo {volume} {467}},\ \bibinfo
		{pages} {459} (\bibinfo {year} {2011})}\BibitemShut {NoStop}%
	\bibitem [{\citenamefont {Bremner}\ \emph {et~al.}(2016)\citenamefont
		{Bremner}, \citenamefont {Montanaro},\ and\ \citenamefont
		{Shepherd}}]{bremner2016averagecase}%
	\BibitemOpen
	\bibfield  {author} {\bibinfo {author} {\bibfnamefont {M.~J.}\ \bibnamefont
			{Bremner}}, \bibinfo {author} {\bibfnamefont {A.}~\bibnamefont {Montanaro}},\
		and\ \bibinfo {author} {\bibfnamefont {D.~J.}\ \bibnamefont {Shepherd}},\
	}\bibfield  {title} {\bibinfo {title} {Average-case complexity versus
			approximate simulation of commuting quantum computations},\ }\href
	{https://doi.org/10.1103/PhysRevLett.117.080501} {\bibfield  {journal}
		{\bibinfo  {journal} {Phys. Rev. Lett.}\ }\textbf {\bibinfo {volume} {117}},\
		\bibinfo {pages} {080501} (\bibinfo {year} {2016})}\BibitemShut {NoStop}%
	\bibitem [{\citenamefont {Leone}\ \emph {et~al.}(2024)\citenamefont {Leone},
		\citenamefont {Oliviero}, \citenamefont {Lloyd},\ and\ \citenamefont
		{Hamma}}]{leone2024learning}%
	\BibitemOpen
	\bibfield  {author} {\bibinfo {author} {\bibfnamefont {L.}~\bibnamefont
			{Leone}}, \bibinfo {author} {\bibfnamefont {S.~F.~E.}\ \bibnamefont
			{Oliviero}}, \bibinfo {author} {\bibfnamefont {S.}~\bibnamefont {Lloyd}},\
		and\ \bibinfo {author} {\bibfnamefont {A.}~\bibnamefont {Hamma}},\ }\bibfield
	{title} {\bibinfo {title} {Learning efficient decoders for quasichaotic
			quantum scramblers},\ }\href {https://doi.org/10.1103/PhysRevA.109.022429}
	{\bibfield  {journal} {\bibinfo  {journal} {Phys. Rev. A}\ }\textbf {\bibinfo
			{volume} {109}},\ \bibinfo {pages} {022429} (\bibinfo {year}
		{2024})}\BibitemShut {NoStop}%
	\bibitem [{\citenamefont {Haug}\ \emph {et~al.}(2025)\citenamefont {Haug},
		\citenamefont {Aolita},\ and\ \citenamefont {Kim}}]{haug2024probing}%
	\BibitemOpen
	\bibfield  {author} {\bibinfo {author} {\bibfnamefont {T.}~\bibnamefont
			{Haug}}, \bibinfo {author} {\bibfnamefont {L.}~\bibnamefont {Aolita}},\ and\
		\bibinfo {author} {\bibfnamefont {M.}~\bibnamefont {Kim}},\ }\bibfield
	{title} {\bibinfo {title} {Probing quantum complexity via universal
			saturation of stabilizer entropies},\ }\href
	{https://doi.org/10.22331/q-2025-07-21-1801} {\bibfield  {journal} {\bibinfo
			{journal} {{Quantum}}\ }\textbf {\bibinfo {volume} {9}},\ \bibinfo {pages}
		{1801} (\bibinfo {year} {2025})}\BibitemShut {NoStop}%
	\bibitem [{\citenamefont {Zhang}\ and\ \citenamefont
		{Zhang}(2025)}]{zhang2025classical}%
	\BibitemOpen
	\bibfield  {author} {\bibinfo {author} {\bibfnamefont {Y.}~\bibnamefont
			{Zhang}}\ and\ \bibinfo {author} {\bibfnamefont {Y.}~\bibnamefont {Zhang}},\
	}\bibfield  {title} {\bibinfo {title} {Classical simulability of quantum
			circuits with shallow magic depth},\ }\href
	{https://doi.org/10.1103/PRXQuantum.6.010337} {\bibfield  {journal} {\bibinfo
			{journal} {PRX Quantum}\ }\textbf {\bibinfo {volume} {6}},\ \bibinfo {pages}
		{010337} (\bibinfo {year} {2025})}\BibitemShut {NoStop}%
	\bibitem [{\citenamefont {Haegeman}\ \emph {et~al.}(2011)\citenamefont
		{Haegeman}, \citenamefont {Cirac}, \citenamefont {Osborne}, \citenamefont
		{Pi\ifmmode~\check{z}\else \v{z}\fi{}orn}, \citenamefont {Verschelde},\ and\
		\citenamefont {Verstraete}}]{haegeman2011timedependent}%
	\BibitemOpen
	\bibfield  {author} {\bibinfo {author} {\bibfnamefont {J.}~\bibnamefont
			{Haegeman}}, \bibinfo {author} {\bibfnamefont {J.~I.}\ \bibnamefont {Cirac}},
		\bibinfo {author} {\bibfnamefont {T.~J.}\ \bibnamefont {Osborne}}, \bibinfo
		{author} {\bibfnamefont {I.}~\bibnamefont {Pi\ifmmode~\check{z}\else
				\v{z}\fi{}orn}}, \bibinfo {author} {\bibfnamefont {H.}~\bibnamefont
			{Verschelde}},\ and\ \bibinfo {author} {\bibfnamefont {F.}~\bibnamefont
			{Verstraete}},\ }\bibfield  {title} {\bibinfo {title} {Time-dependent
			variational principle for quantum lattices},\ }\href
	{https://doi.org/10.1103/PhysRevLett.107.070601} {\bibfield  {journal}
		{\bibinfo  {journal} {Phys. Rev. Lett.}\ }\textbf {\bibinfo {volume} {107}},\
		\bibinfo {pages} {070601} (\bibinfo {year} {2011})}\BibitemShut {NoStop}%
	\bibitem [{\citenamefont {Valiant}(2001)}]{valiant2001quantumcomputers}%
	\BibitemOpen
	\bibfield  {author} {\bibinfo {author} {\bibfnamefont {L.~G.}\ \bibnamefont
			{Valiant}},\ }\bibfield  {title} {\bibinfo {title} {Quantum computers that
			can be simulated classically in polynomial time},\ }in\ \href
	{https://doi.org/10.1145/380752.380785} {\emph {\bibinfo {booktitle}
			{Proceedings of the Thirty-Third Annual ACM Symposium on Theory of
				Computing}}},\ \bibinfo {series and number} {STOC '01}\ (\bibinfo
	{publisher} {Association for Computing Machinery},\ \bibinfo {address} {New
		York, NY, USA},\ \bibinfo {year} {2001})\ p.\ \bibinfo {pages}
	{114–123}\BibitemShut {NoStop}%
	\bibitem [{\citenamefont {Wu}\ \emph {et~al.}(2025)\citenamefont {Wu},
		\citenamefont {Kloss}, \citenamefont {Krinitsin}, \citenamefont {Fishman},
		\citenamefont {Pixley},\ and\ \citenamefont
		{Stoudenmire}}]{wu2025disentangling}%
	\BibitemOpen
	\bibfield  {author} {\bibinfo {author} {\bibfnamefont {A.-K.}\ \bibnamefont
			{Wu}}, \bibinfo {author} {\bibfnamefont {B.}~\bibnamefont {Kloss}}, \bibinfo
		{author} {\bibfnamefont {W.}~\bibnamefont {Krinitsin}}, \bibinfo {author}
		{\bibfnamefont {M.~T.}\ \bibnamefont {Fishman}}, \bibinfo {author}
		{\bibfnamefont {J.~H.}\ \bibnamefont {Pixley}},\ and\ \bibinfo {author}
		{\bibfnamefont {E.~M.}\ \bibnamefont {Stoudenmire}},\ }\bibfield  {title}
	{\bibinfo {title} {Disentangling interacting systems with fermionic
			{Gaussian} circuits: Application to quantum impurity models},\ }\href
	{https://doi.org/10.1103/PhysRevB.111.035119} {\bibfield  {journal} {\bibinfo
			{journal} {Phys. Rev. B}\ }\textbf {\bibinfo {volume} {111}},\ \bibinfo
		{pages} {035119} (\bibinfo {year} {2025})}\BibitemShut {NoStop}%
	\bibitem [{\citenamefont {Hebenstreit}\ \emph {et~al.}(2019)\citenamefont
		{Hebenstreit}, \citenamefont {Jozsa}, \citenamefont {Kraus}, \citenamefont
		{Strelchuk},\ and\ \citenamefont
		{Yoganathan}}]{hebenstreit2019allpurefermionic}%
	\BibitemOpen
	\bibfield  {author} {\bibinfo {author} {\bibfnamefont {M.}~\bibnamefont
			{Hebenstreit}}, \bibinfo {author} {\bibfnamefont {R.}~\bibnamefont {Jozsa}},
		\bibinfo {author} {\bibfnamefont {B.}~\bibnamefont {Kraus}}, \bibinfo
		{author} {\bibfnamefont {S.}~\bibnamefont {Strelchuk}},\ and\ \bibinfo
		{author} {\bibfnamefont {M.}~\bibnamefont {Yoganathan}},\ }\bibfield  {title}
	{\bibinfo {title} {All pure fermionic non-{Gaussian} states are magic states
			for matchgate computations},\ }\href
	{https://doi.org/10.1103/PhysRevLett.123.080503} {\bibfield  {journal}
		{\bibinfo  {journal} {Phys. Rev. Lett.}\ }\textbf {\bibinfo {volume} {123}},\
		\bibinfo {pages} {080503} (\bibinfo {year} {2019})}\BibitemShut {NoStop}%
\end{thebibliography}
\end{document}